\documentclass{article}

\usepackage{arxiv}

\usepackage[utf8]{inputenc} 
\usepackage[T1]{fontenc}    
\usepackage{hyperref}       
\usepackage{url}            
\usepackage{booktabs}       
\usepackage{amsfonts}       
\usepackage{nicefrac}       
\usepackage{microtype}      
\usepackage{lipsum}
\usepackage{graphicx}
\graphicspath{ {./images/} }
\usepackage{amsmath}
\usepackage{multirow}
\usepackage{makecell}
\setcellgapes{5pt}
\usepackage{float}
\usepackage{xcolor}

\usepackage[toc,page]{appendix}

\newcommand{\revised}[1]{\textcolor{black}{#1}}

\makeatletter
\newcommand*{\rom}[1]{\expandafter\@slowromancap\romannumeral #1@}
\makeatother

\title{Generative Inverse Design of Metamaterials with Functional Responses by Interpretable Learning}

\author{
 Wei (Wayne) Chen \\
  J. Mike Walker ’66 Department of Mechanical Engineering\\
  Texas A\&M University\\
  College Station, TX 77843 \\
  \texttt{w.chen@tamu.edu} \\
   \And
 Rachel Sun \\
  Department of Mechanical Engineering\\
  Massachusetts Institute of Technology\\
  Cambridge, MA 02139 \\
  \texttt{rmsun@mit.edu} \\
  \And
 Doksoo Lee \\
  Department of Mechanical Engineering\\
  Northwestern University\\
  Evanston, IL 60208 \\
  \texttt{dslee@northwestern.edu} \\
  \And
 Carlos M. Portela \\
  Department of Mechanical Engineering\\
  Massachusetts Institute of Technology\\
  Cambridge, MA 02139 \\
  \texttt{cportela@mit.edu} \\
  \And
 Wei Chen \\
  Department of Mechanical Engineering\\
  Northwestern University\\
  Evanston, IL 60208 \\
  \texttt{weichen@northwestern.edu} \\
}

\begin{document}
\maketitle
\begin{abstract}
Metamaterials with functional responses can exhibit varying properties under different conditions (e.g., wave-based responses or deformation-induced property variation). This work addresses the rapid inverse design of such metamaterials to meet target qualitative functional behaviors, a challenge due to its intractability and non-unique solutions. Unlike data-intensive and non-interpretable deep-learning-based methods, we propose the Random-forest-based Interpretable Generative Inverse Design (RIGID), a single-shot inverse design method for fast generation of metamaterial designs with on-demand functional behaviors. RIGID leverages the interpretability of a random forest-based ``design$\rightarrow$response'' forward model, eliminating the need for a more complex ``response$\rightarrow$design'' inverse model. Based on the likelihood of target satisfaction derived from the trained random forest, one can sample a desired number of design solutions using Markov chain Monte Carlo methods. We validate RIGID on acoustic and optical metamaterial design problems, each with fewer than 250 training samples. 
\revised{Compared to the genetic algorithm-based design generation approach, RIGID generates satisfactory solutions that cover a broader range of the design space, allowing for better consideration of additional figures of merit beyond target satisfaction.}
This work offers a new perspective on solving on-demand inverse design problems, showcasing the potential for incorporating interpretable machine learning into generative design under small data constraints.
\end{abstract}


\section{Introduction}
\label{sec:introduction}

\revised{
Metamaterials are engineered materials, typically composed of periodically arranged building blocks (i.e., unit cells), that exhibit properties and functionalities beyond their constituent materials~\cite{kadic2013metamaterials,bertoldi2017flexible}. We can achieve certain effective material properties of metamaterials by designing their structure (i.e., geometries of unit cells) rather than chemical composition. 
Optical~\cite{soukoulis2011past}, acoustic~\cite{cummer2016controlling}, thermal~\cite{schittny2013experiments}, and mechanical metamaterials~\cite{christensen2015vibrant,bertoldi2017flexible} are metamaterials with unique properties tailored to manipulate specific types of waves or energy. For example, optical metamaterials or metasurfaces (i.e., the two-dimensional form of metamaterials) are designed to control and manipulate electromagnetic waves, which can lead to unusual properties such as a negative refractive index~\cite{pendry2000negative}. Acoustic metamaterials are designed to control how sound (acoustic) waves propagate which enables functionalities such as noise reduction~\cite{gao2022acoustic}.
}

\revised{Functional responses refer to} varying properties or behaviors under different conditions. A notable example that involves functional responses is metamaterials, which alter their electromagnetic, acoustic, or elastic wave propagation behaviors depending on the wavelength or frequency~\cite{jiang2022flexible}. Another example is metamaterials that exhibit changing properties or functionalities due to deformation in response to external stimuli like temperature~\cite{boley2019shape} and magnetic fields~\cite{ma2022deep}.
Tailoring functional responses of these metamaterials is of interest to applications such as sound and vibration control, analog computing, medical imaging, sensing, communication, and soft robotics. In many use cases, rather than precisely controlling the complete functional responses, we only care about \textit{qualitative} behaviors under certain conditions. For example, acoustic metamaterials were usually designed to have bandgaps at specified frequencies to achieve functionalities like wave-guiding~\cite{casadei_piezoelectric_2012,liu2020acoustic}, focusing~\cite{kim2021poroelastic,xie2018acoustic}, and vibration mitigation~\cite{matlack2016composite,krodel2015wide,bayat2018wave}. However, it is unnecessary and computationally expensive to design for the whole dispersion relation~\cite{ronellenfitsch2019inverse,goh2020inverse,li2020designing,zhang2021realization,chen2022see}. Similarly, we may design optical metamaterials to qualitatively manipulate optical properties (e.g., high or low absorption/reflection/transmission) under certain wavelengths, without requiring the entire spectral response to match an exact target~\cite{vismara2019solar,cheng2021plasmonic}.

Identifying metamaterial designs from a given target forms an inverse design problem. Unlike many forward problems where one can obtain solutions (e.g., spectral responses or material properties under external stimuli) by modeling the physics or conducting experiments, inverse design problems are usually intractable. Traditionally, these problems are solved by iterative optimization (i.e., minimizing the difference between the actual quantity of interest and the target)~\cite{ronellenfitsch2019inverse,goh2020inverse,liu2020compounding,zhang2021realization,Lee2024DeepNeuralOperator}. This, however, requires repeatedly updating the design solution and solving forward problems. When the design target changes, one needs to rerun the entire optimization process. Thus, inverse design by iterative optimization becomes impractical if solving the forward problem (by simulations or experiments) is time-consuming or if the design target needs to change frequently. For example, in the case of acoustic metamaterial design, one may want to obtain a material waveguide containing spatially-varying unit cells with different bandgap ranges at different locations in a tessellation. To accelerate the optimization approach, prior works replaced simulations or experiments with machine learning models~\cite{inampudi2018neural,zhelyeznyakov2021deep}. However, the efficiency and quality of final solutions are highly dependent on both the machine learning model and the optimization algorithm. On the other hand, a single run of optimization usually only returns one final solution, although multiple designs might satisfy a given target (i.e., the \textit{non-uniqueness} of solutions). For example, multiple acoustic metamaterial designs may have bandgaps within the same target frequency range. This non-uniqueness nature of inverse design problems was also shown for optical metasurfaces~\cite{liu2018training, ma2019probabilistic, An2021MultifunctionalNetwork,Lee2024DeepNeuralOperator}, especially when we only care about qualitative target responses. \revised{It is generally advantageous for inverse design approaches to yield a diverse set of solutions, as this allows consideration of additional figures of merit beyond target satisfaction. For example, deriving diverse solutions for acoustic metamaterials will allow us to consider the kinematic compatibility of multiple designs in waveguide design problems.} However, optimization approaches cannot explore diverse alternative solutions efficiently. 

In this work, we aim to achieve iteration-free, single-shot inverse design for metamaterials~\textemdash~given a qualitative target response, we want to generate multiple satisfying design solutions rapidly, without the need for iteratively evaluating designs (i.e., running simulations or conducting experiments); and after one-time initial training, no additional training or design evaluation is required to generate designs for different targets. This allows fast and comprehensive exploration of the feasible design space under different target responses.

Prior research attempted to realize iteration-free, single-shot inverse design using machine learning. There are three mainstream methods. The most straightforward approach is to learn a \textit{direct inverse mapping} from the response to design variables. Neural networks are the most commonly used machine learning model for this purpose, due to their high flexibility in approximating arbitrary nonlinear input-output relationships~\cite{Malkiel2018PlasmonicLearning,li2020designing}. Despite the simplicity of the direct inverse mapping, its underlying assumption of the response-design mapping being one-to-one does not hold in many cases due to the non-uniqueness of solutions, as mentioned earlier. Such non-uniqueness will cause conflicting training instances where the same input (response) is associated with distinct outputs (designs), which will destabilize the convergence during neural network training~\cite{liu2018training,lee2023data}. To avoid this issue, past work proposed the \textit{Tandem Neural Network (T-NN)} that cascades an inverse-design network with a pretrained forward-modeling network to avoid using designs as labels when training the inverse-design network and hence solved the training convergence issue~\cite{liu2018training,an2019deep,kumar2020inverse,Yeung2021MultiplexedNetworks,bastek2022inverting,xie2023deep,mahesh2024deep}.
Nonetheless, the original T-NNs still learn a one-to-one response-design mapping and cannot account for the non-uniqueness of design solutions. To fundamentally solve this problem, one needs to learn a one-to-many mapping. For example, Bastek et al.~\cite{bastek2022inverting} integrated stochastic sampling into the inverse-design network to allow the generation of multiple feasible solutions. Ha et al.~\cite{ha2023rapid} used multiple T-NNs to predict multiple design solutions given a target response. Wang et al.~\cite{wang2023demand} adapted the T-NN to mimic the structure of a variational autoencoder that is capable of learning the conditional distributions of designs given target responses. A large body of recent works achieved the goal of learning one-to-many mapping by using \textit{conditional generative models}, typically conditional generative adversarial networks (cGANs)~\cite{jiang2019free,so2019designing,wen2020robust,gurbuz2021generative,An2021MultifunctionalNetwork}, conditional variational autoencoders (cVAEs)~\cite{ma2019probabilistic,ma2022pushing,lei2024dynamic}, and conditional diffusion models~\cite{lew2023single,zhang2023diffusion,bastek2023inverse}. These models can generate multiple designs given a target response by learning the distribution of designs conditioned on the response. 
We include more detailed explanations of these three mainstream methods in \textit{SI Appendix, Sec.~S1 and Fig.~S1}. 

We refer interested readers to Ref.~\cite{lee2023data,so2023revisiting,zheng2023deep,jin2022intelligent,ma2021deep,jiang2021deep,xu2023software} for comprehensive reviews of existing metamaterial design methods. 
To the best of our knowledge, almost all the existing iteration-free, single-shot metamaterial inverse design methods are based on deep learning, which has many common issues
such as high data demand, exhaustive hyperparameter tuning, slow training, and low interpretability, especially compared to traditional machine learning models like decision trees and random forests. 
On the other hand, Elzouka et al.~\cite{elzouka2020interpretable} proposed to use the decision tree as a more interpretable model to solve both the forward prediction and inverse design problem. After training a decision tree for forward prediction, one can identify explicit design rules (i.e., feasible regions in the design space) by tracing from target leaf nodes to the root node. This approach also captures the one-to-many mapping nature of inverse design problems since it gives feasible design variable ranges rather than a single solution. However, there remain some limitations. Firstly, for solutions identified by the design rules, the method does not differentiate their likelihood of target satisfaction. Yet in reality, solutions always have different likelihoods due to the uncertainty of model estimation. Secondly, the method has to train two models: a random forest was trained first to ensure model accuracy and robustness, and then a large decision tree was trained to emulate the performance of the random forest and provide design rules. This is due to the challenge of deriving explicit design rules from an ensemble model like the random forest. Finally, the method was demonstrated on a problem with more than $10^4$ training data, while the effectiveness on smaller datasets (i.e., data with orders of magnitude smaller sample sizes) was not studied.

\begin{figure}
\centering
\includegraphics[width=\textwidth]{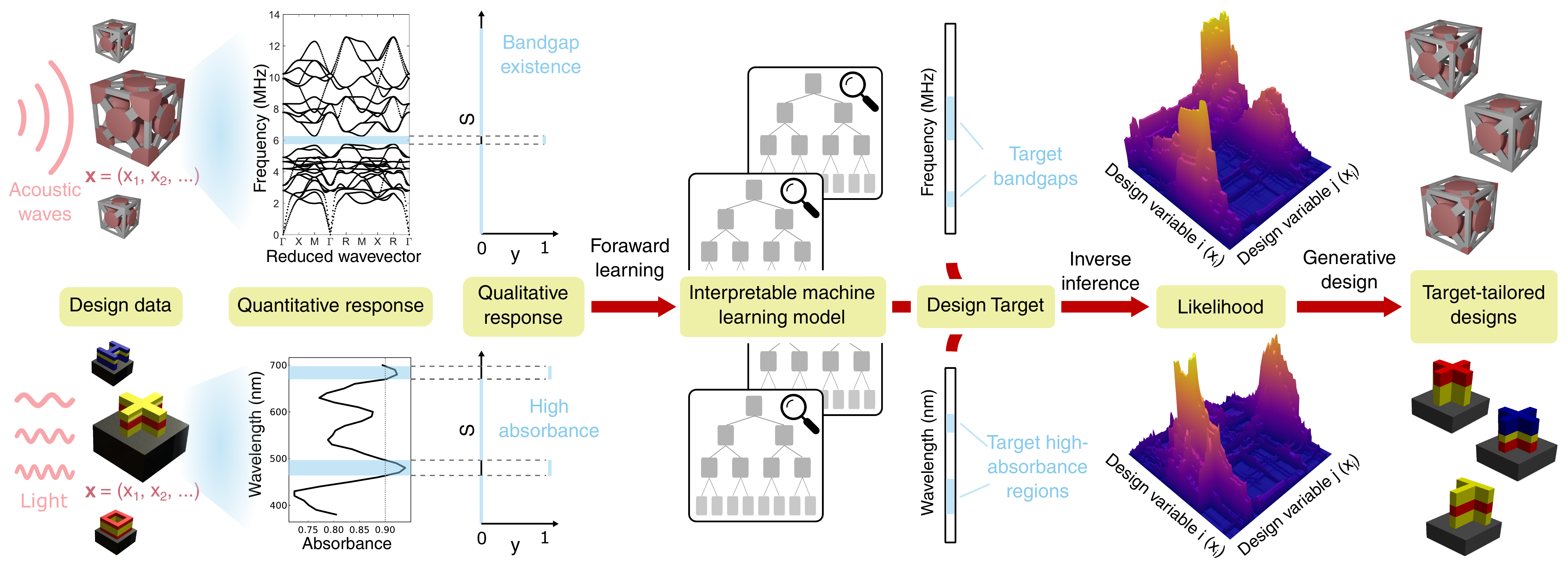}
\caption{Schematic diagram of the RIGID method. We first train a random forest on a design-response dataset to learn the forward design-response relation~\textemdash~predicting qualitative responses (e.g., bandgap existence at any given wave frequency) of designs. Then given a design target, we can infer the likelihood of any design satisfying the target by probing into the trained random forest. New designs with tailored responses can be generated by sampling the design space based on the likelihood estimation.}
\label{fig:overview}
\end{figure}

This work aims to address the aforementioned problems by proposing a method called \textit{Random-forest-based Interpretable Generative Inverse Design (RIGID)}. 
Figure~\ref{fig:overview} shows an overview of this method. Specifically, we first train a forward prediction random forest. Then given a design target, we can probe the trained random forest to infer the likelihood of any design satisfying the target. To generate new designs tailored to the target, we can sample from the design space according to the likelihood. Compared to the most widely studied neural-network-based methods, RIGID has a much lower cost in training and hyperparameter tuning, and works more robustly on small-size datasets (as random forests are less prone to overfitting). Similar to deep generative models, it can generate a desired number of solutions, allowing one to explore alternative solutions that might have desired properties or functionalities beyond the ones considered as the target. The explicit likelihood estimation also offers an interpretable characterization of a design's target satisfaction probability and allows an exploitation-exploration trade-off when selecting generated designs. We validate the RIGID method on two metamaterial design examples~\textemdash~an acoustic metamaterial design example, where the target is to generate metamaterials with specific bandgaps, and an optical metasurface design example, where the target is to generate metasurfaces with high absorbance at specified wavelengths.

Our contributions are three-fold. First, we propose an iteration-free, single-shot inverse design method that is fast, generative, interpretable, and small-data-compatible. Secondly, we demonstrate the effectiveness of the proposed method on acoustic and optical metamaterial design examples, and propose both qualitative and quantitative ways of assessing our method. Finally, we create two synthetic test cases for fast examination and validation of model performance. These test cases can be used for future benchmarking studies of related methods.




\section{Methods}
\label{sec:methodology}

The functional response of metamaterials can be modeled as $y = f(\mathbf{x}, s)$, where $\mathbf{x}$ denotes metamaterial design variables (e.g., materials and geometry parameters), $s$ is an \textit{auxiliary variable} representing the independent variable (or the ``$x$-axis'') of the response (e.g., the frequency/wavelength or the external stimuli such as temperature), and $y$ indicates the value of the response associated with our design target. In this paper, we assume $y \in \{0, 1\}$ since we only focus on qualitative behaviors at specified frequencies (e.g., for an acoustic metamaterial or an optical metamaterial design, whether a bandgap exists or whether the energy absorbance is higher than a threshold within a range of frequencies). We leave the more challenging problem of tailoring quantitative behaviors as future work.

Our goal is to solve the inverse problem~\textemdash~find a set of design solutions that satisfy $f(\mathbf{x}, s) = 1$ when $s$ is inside a target domain $\Omega$, i.e., $\{\mathbf{x}^*|f(\mathbf{x}^*, s) = 1, \forall s \in \Omega\}$. Here $f(\mathbf{x}, s) = 1$ could mean, for instance, the existence of bandgaps for acoustic metamaterials or high absorption for optical metamaterials. The domain $\Omega$ can be any frequency or wavelength interval(s). Note that this design target only requires the satisfaction of behaviors within the specified domain $\Omega$, while behaviors outside this domain are out of concern in this study, although it is straightforward to adopt our method to this problem setting.

We use a random forest to approximate the function $f$. A random forest is an ensemble learning method that combines the predictions of multiple decision trees to improve accuracy and reduce overfitting~\cite{breiman2001random}. The trained random forest serves as a \textit{forward prediction} model that predicts the outcome $y$ given design variables $\mathbf{x}$ and the auxiliary variable $s$. Compared to the widely used neural networks, the random forest as a forward prediction model (1)~offers significantly faster training, (2)~has fewer hyperparameters to tune, (3)~is less susceptible to overfitting on small data, and (4)~has higher interpretability (i.e., the decision-making of each tree in the random forest is transparent). More importantly, this interpretability also allows us to realize \textit{inverse design} without training a separate inverse model.

\begin{figure}[h]
\centering
\includegraphics[width=.9\textwidth]{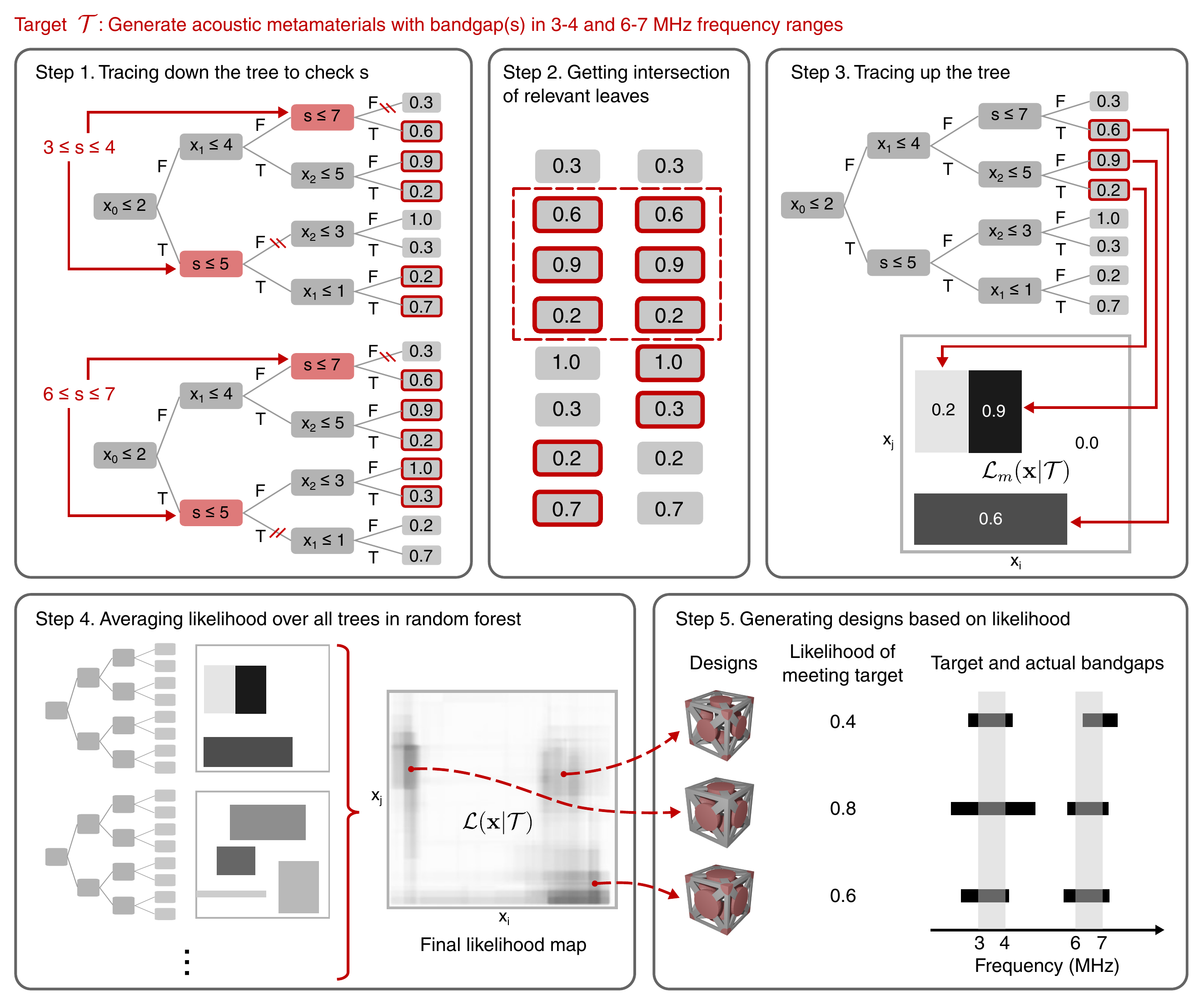}
\caption{The inverse design pipeline of the proposed RIGID method (using the inverse design of acoustic metamaterials as an example). Given design parameters $\mathbf{x}$ and the auxiliary variable $s$ (e.g., wave frequency), a trained random forest predicts the probability of the qualitative response $y$ (e.g., bandgap existence). Each tree in the random forest splits the joint space of $\mathbf{x}$ and $s$ into regions, each associated with a specific prediction (shown on leaf nodes). The splitting criteria are encoded in tree nodes. ``T'' means meeting a criterion and ``F'' means not meeting it. RIGID first identifies leaf nodes that are relevant to the considered range of auxiliary variable $s$ by checking splitting criteria related to $s$ and pruning tree branches that are irrelevant (Step 1). If the considered range of $s$ has multiple parts, we repeat this step for each part and take the intersection of relevant leaves (Step 2). Each relevant leaf node corresponds to a decision path indicating a region in the design space, as well as a predicted probability of target satisfaction, which is a score we assign to the corresponding design space region (Step 3). With multiple trees in a random forest, we can average the scores predicted by each tree and use the average score as our likelihood estimation (Step 4). We can then sample from the design space based on the likelihood distribution to generate new designs tailored to the target (Step 5). Note that the 2-dimensional likelihood maps are only for visualization purposes. The actual dimension will be the same as the design space dimension (i.e., the number of design variables).}
\label{fig:rigid}
\end{figure}

Figure~\ref{fig:rigid} shows how, by probing the trained random forest, one can estimate a likelihood distribution for target satisfaction of solutions over the entire design space and sample (generate) new designs based on this likelihood distribution. Since we target qualitative (binary) behaviors at specified $s$ (e.g., a bandgap in 3-4 MHz frequency or high absorption at a wavelength of 400-500 nm), we first identify the leaf nodes (on each decision tree in the random forest) that are relevant to the $s$ in the target (Fig.~\ref{fig:rigid}, Step 1). We do this by tracing down each tree, checking only the nodes that use $s$ as the splitting feature, and pruning the branches that are irrelevant to the $s$ in the target. For example, as shown in Fig.~\ref{fig:rigid}, there are two tree nodes using $s$ as the splitting feature, with splitting criteria at $s \leq 5$ and $s \leq 7$. Given the target frequency range of $3 \leq s \leq 4$, we can remove the right branches of both nodes as these branches are only relevant to $s>5$ and $s>7$, respectively, which conflicts with the target range of $3 \leq s \leq 4$. After pruning these branches, we end up with a set of leaves relevant to the target (highlighted in Fig.~\ref{fig:rigid}, Step 1). When we have a combined target (e.g., bandgaps in both 3-4 MHz and 6-7 MHz, as shown in Fig.~\ref{fig:rigid}), we need to get the intersection of all the sets of relevant leaves and use that as the final set of relevant leaves (highlighted in Fig.~\ref{fig:rigid}, Step 2). Note that a combined target includes cases where there are multiple nonadjacent target ranges (e.g., 3-4 MHz and 6-7 MHz) or when a target range is split by a tree node (e.g., a target range of 4-6 MHz can be split by the node ``$s \leq 5$'', thus we need to consider it as the combination of two target ranges~\textemdash~4-5 MHz and 5-6 MHz). A more detailed discussion of this step is in \textit{SI Appendix, Sec.~S2}.

The next step is to trace up the tree from the $N$ relevant leaves, obtained by Step 2, to the root node (Fig.~\ref{fig:rigid}, Step 3). This will result in $N$ decision paths, along which are nodes indicating splitting criteria for design variables $\mathbf{x}$. Thus, each decision path represents a set of design variable ranges, or in other words, a region in the design space. We assign each region a score equal to the predicted probability at each corresponding leaf. This probability is learned from the training data and equals the proportion of positive data in a leaf. It indicates the tree's belief in the probability of a design $\mathbf{x}$ satisfying the target $\mathcal{T} = \{f(\mathbf{x}^*, s) = 1 | \forall s \in \Omega\}$ ($\Omega=[3,4] \cup [6,7]$ in this example) if the design falls in the design space region corresponding to the leaf. Therefore, with a single decision tree $m$, we already have the map of likelihood $\mathcal{L}_m(\mathbf{x}|\mathcal{T})=\mathbb{P}_m(\mathcal{T}|\mathbf{x})$ for target satisfaction: each of the $N$ regions has a uniformly distributed likelihood equal to the predicted probability at the corresponding leaf, and the rest of the design space has a likelihood of 0 (Fig.~\ref{fig:rigid}, Step 3).

Since a single decision tree usually lacks accuracy, robustness, and a way to quantify estimation uncertainty, we still want to take advantage of the random forest as an ensemble model for inverse design. We use Steps 1-3 to derive the likelihood distribution for each of the $M$ trees in the random forest, and simply use the average of these $M$ likelihood distributions as the final likelihood for target satisfaction, $\mathcal{L}(\mathbf{x}|\mathcal{T})=\sum_{m=1}^M \mathcal{L}_m(\mathbf{x}|\mathcal{T})/M$, which is a more complex and smooth function (Fig.~\ref{fig:rigid}, Step 4). If more trees believe a design $\mathbf{x}$ has a higher likelihood of satisfying the target, then the design will have a higher likelihood $\mathcal{L}(\mathbf{x}|\mathcal{T})$. Finally, to generate new designs, we can sample from $\mathcal{L}(\mathbf{x}|\mathcal{T})$ using Markov chain Monte Carlo (MCMC) methods such as Metropolis-Hastings~\cite{hastings1970monte} (Fig.~\ref{fig:rigid}, Step 5).

\revised{
We can also derive the posterior of a design $\mathbf{x}$ given the target if we know the prior $p(\mathbf{x})$ based on Bayes' theorem:
$$p(\mathbf{x}|\mathcal{T}) \propto \mathcal{L}(\mathbf{x}|\mathcal{T})p(\mathbf{x})$$
In this work, we assume a uniform prior, making the posterior directly proportional to the likelihood $\mathcal{L}(\mathbf{x}|\mathcal{T})$, and sampling from the posterior is equivalent to sampling from the likelihood.
}

Compared to prior works, RIGID provides the following unique benefits:
\begin{enumerate}
    \item It is effective on small data problems as the random forest is less susceptible to overfitting.
    \item The training is fast (in seconds of wall time) and does not require computationally-demanding hyperparameter tuning. Once the training is done, no further training or iterative optimization is required to generate designs for different targets.
    \item The method estimates the explicit likelihood of target satisfaction for every possible solution in the design space. Given a design target of specific functional behavior, we can generate an unlimited number of solutions based on the likelihood, allowing us to explore alternative solutions that might have desired properties or functionalities beyond the ones considered as the target. 
    \item The method offers a high level of transparency as one can easily probe the trained model to understand its reasoning behind any decision-making (i.e., why a design has a high/low likelihood).
    \item When generating design solutions, one can use a single parameter~\textemdash~the sampling threshold~\textemdash~to easily tune the trade-off between exploitation (i.e., generated designs have higher chances of satisfying the target) and exploration (i.e., generated designs cover a broader area of the design space), as we will demonstrate in Results.
\end{enumerate}

\section{Results}
\label{sec:results}

We demonstrate our RIGID method on an acoustic metamaterial design problem, an optical metasurface design problem, and two synthetic design problems. Based on a recent review article by Lee et al.~\cite{lee2023data} and other related works (e.g., \cite{bastek2022inverting}), existing iteration-free, single-shot inverse design methods were demonstrated on training data size ranging from $10^3$ to $10^6$ in scale. Here we show that our method can work with much smaller-scale datasets (less than 250 training samples).

For all the test problems, we used the same random forest hyperparameter settings and did not perform hyperparameter tuning. Specifically, each random forest contains 1,000 trees with a minimum of 2 samples required to split an internal node and a minimum of 1 sample required to be at a leaf node. Gini impurity~\cite{BreiFrieStonOlsh84} was used as the splitting criterion at tree nodes.
The train-test split ratio was 4:1. Since the positive/negative training data can be highly imbalanced (e.g., the frequency ranges with bandgaps are much narrower than those without), we used the Synthetic Minority Over-sampling TEchnique (SMOTE)~\cite{chawla2002smote} to over-sample the positive class. For all the case studies, the random forest training took less than 10 seconds on an Intel Core i5-9300H CPU 2.4GHz and 8GB memory. After training, we generate new designs by sampling from the resulting likelihood distribution using Metropolis-Hastings. In practice, Metropolis-Hastings can generate identical samples, which provides no benefits for design purposes. Thus in this work, we reject the designs identical to previous ones during sampling.

\revised{We compare RIGID to a commonly used low-cost generative design strategy\textemdash surrogate modeling combined with a genetic algorithm (GA)~\cite{liu2020hybrid,garland2021pragmatic, zhang2021genetic,shen2022nature,lee2022generative}. We choose this baseline according to the following criteria to ensure that its efficiency and application scenario are comparable to those of RIGID: 1)~the baseline method should generate design solutions quickly (i.e., no additional training or design evaluation is required given different design targets), 2)~the baseline method can generate multiple satisfying solutions to the inverse design problem, and 3)~the baseline method can handle small data. These criteria exclude sequential design strategies like Bayesian optimization (as they violate the first criterion) and any existing single-shot inverse design methods mentioned in Sec.~\ref{sec:introduction} (as direct inverse mapping and T-NN violate the second criterion and conditional generative model-based methods usually violate the last criterion).}

\revised{Genetic algorithms are optimization methods that iteratively evolve a population of candidate solutions, based on biologically inspired operators such as mutation, crossover, and selection, toward better solutions~\cite{mitchell1998introduction}. As we usually need to evaluate a large population of candidate solutions at each iteration (referred to as a \textit{generation}), a surrogate model (usually a machine learning model) is often constructed to accelerate design response prediction, thereby speeding up the optimization process. For the surrogate model in the baseline method, we use the same trained random forest as in RIGID for consistency, eliminating differences due to forward prediction accuracy and focusing on comparing the inverse design methods. We set the objective of GA to maximize the aggregated probability of target satisfaction within the target domain $\Omega$, i.e., $\max_{\mathbf{x}} \sum_{s_i\in\Omega} f'(\mathbf{x}, s_i)$, where $\{s_i\}_{i=1}^n$ are discretized auxiliary variable values (e.g., discrete values of frequencies or wavelengths) and $f'$ represents the trained random forest. We include the detailed configurations of GA in Sec.~\ref{sec:experimental_secion}. After optimization, we rank the solutions generated throughout the optimization process based on their objective values and select the top solutions as final generated designs. We compare these designs to those generated by RIGID. From this point, we will call designs generated by RIGID and GA as \textit{RIGID designs} and \textit{GA designs}, respectively.
}

\subsection{Applying RIGID to Design Acoustic Metamaterials with Target Bandgaps}
\label{sec:acoustic}

\begin{figure}[h]
\centering
\includegraphics[width=.8\textwidth]{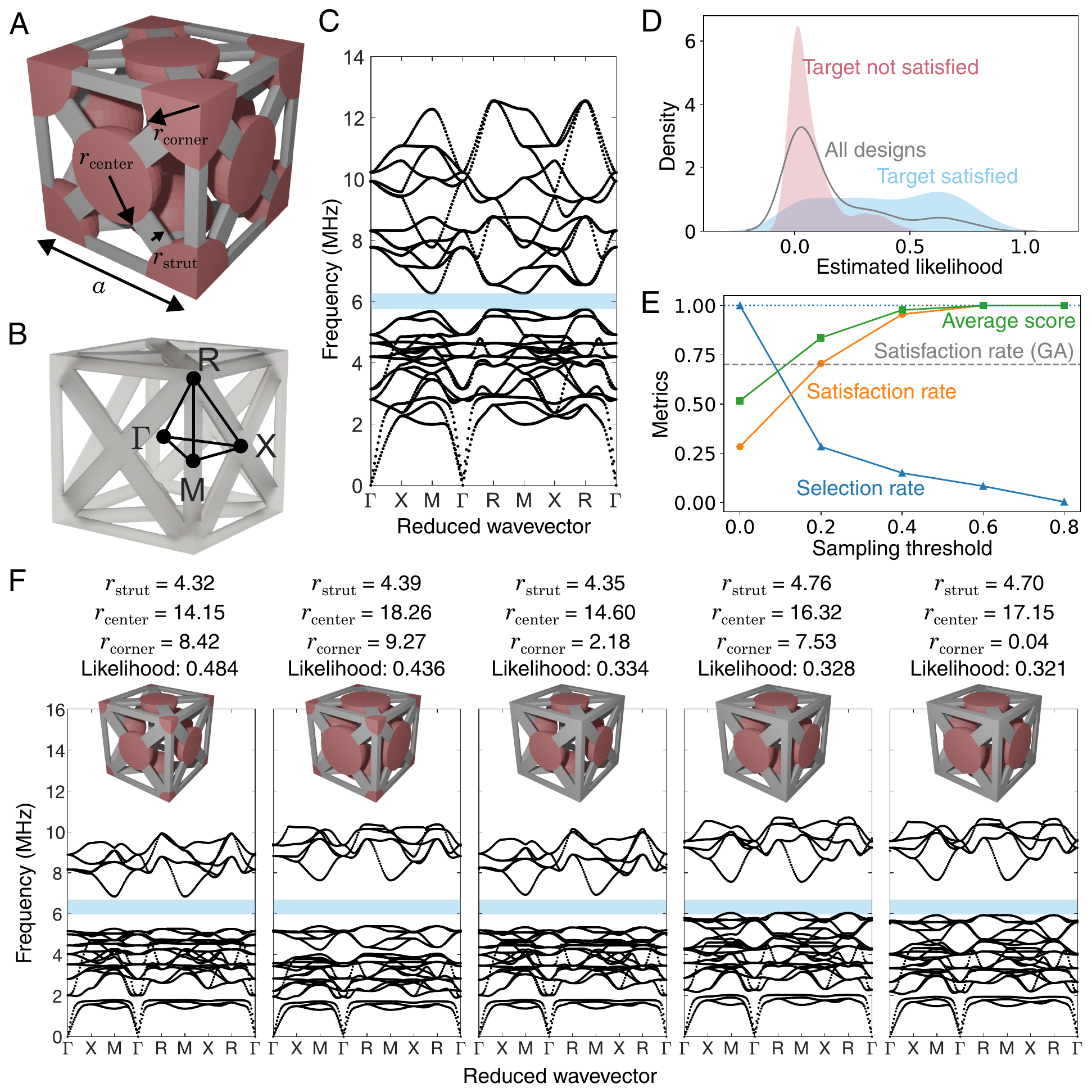}
\caption{Acoustic metamaterial design problem configuration and results. (A)~Design variables of center and corner mass radii ($r_\text{center}$ and $r_\text{corner}$) and strut radius ($r_\text{strut}$). (B)~High symmetry points of the cubic irreducible Brillouin zone. (C)~A sample dispersion relation and bandgap (marked by the highlighted zone). The design objective is to generate new acoustic metamaterial designs with target bandgaps. (D)~KDE of the estimated likelihood for generated designs. \revised{(E)~Satisfaction rates, average scores, and selection rates for RIGID designs under varying sampling thresholds (solid lines), in comparison to the satisfaction rate of GA designs (horizontal dashed line). The horizontal dotted line indicates 100\% satisfaction.} (F)~Geometries and corresponding dispersion relations of five RIGID designs with the highest likelihood of satisfying a specified target bandgap (marked as highlighted frequency regions). All the radii ($r_\text{strut}$, $r_\text{center}$, and $r_\text{corner}$) have a unit of $\mu$m. Here only the fourth design fails to meet a small portion (at around 6 MHz) of the target bandgap, whereas the others meet it. Generated designs for other targets are shown in \textit{SI Appendix, Figs.~S2-S6}.
}
\label{fig:acoustic}
\end{figure}

Here we consider acoustic metamaterials that can control elastic wave propagation at ultrasound (MHz) frequencies. Varying the microscale geometries of acoustic metamaterials changes the dynamic properties of a material, such as bandgaps~\cite{bayat2018wave} (i.e., forbidden frequency ranges of a material) and wave propagation direction~\cite{casadei_piezoelectric_2012}. 
These materials promise applications in waveguides~\cite{casadei_piezoelectric_2012,liu2020acoustic}, lenses~\cite{kim2021poroelastic,xie2018acoustic}, and vibration mitigation~\cite{krodel2015wide}. Designing acoustic metamaterials with target bandgaps is challenging, as many three-dimensional architectures do not naturally have full bandgaps.
We present the braced cubic design framework (Fig. \ref{fig:acoustic}A-B) as a method to tune the size and location of bandgaps (Fig.~\ref{fig:acoustic}C). 
In particular, spherical micro-inertia are added to the center and corner of a braced cubic unit cell with strut radius $r_\text{strut}$. 
Micro-inertia placed at the center of the brace has radius $r_\text{center}$ while micro-inertia placed at the corner of the cubic unit cell has radius $r_\text{corner}$. 

We randomly created 284 sets of geometric parameters $\mathbf{x} = (r_\text{strut}, r_\text{center}, r_\text{corner})$ with 4 $\leq r_\text{strut} \leq$ 6.41, 0 $\leq r_\text{center} \leq$ 20, and 0 $\leq r_\text{corner} \leq$ 20 (unit: \textmu m). The unit cell size was set at $a=60$ \textmu m. For each of these designs, we performed Bloch-wave analysis to compute its acoustic dispersion relation. 
Bandgap location and width were extracted for each design based on its dispersion relation. 

Out of the 284 sets of design variables and bandgap data, we used 227 samples as training data. We first discretized the entire frequency range into 100 intervals, and trained a random forest to predict bandgap existence $y\in\{0,1\}$ at a specific interval $s$ for a given design $\mathbf{x}$. The trained model has a test F1 score of 0.82. The resulting confusion matrix on test data is shown in \textit{SI Appendix, Tab.~S1}.

To test the inverse design capability of RIGID, we randomly created 10 targets, each containing 1-2 frequency ranges in which bandgap(s) should exist. 
We generated 30 designs for each target by sampling from the resulting likelihood distribution over the design space\footnote{Note that it is possible for the likelihood to be zero everywhere in the design space when the model believes the target is unachievable. We ignore these cases as it is meaningless and impossible to sample designs from such likelihood distribution.}. Bandgaps were identified from dispersion relations computed using Bloch-wave analysis.

Figure~\ref{fig:acoustic}D shows the kernel density estimation (KDE) for the likelihood of the 300 RIGID designs, conditioned on their target satisfaction.
We use $\mathcal{D}$ and $\mathcal{D}_\text{feas}$ to represent the complete set of generated designs and the set of generated designs that actually satisfy the target, respectively. Then $\mathcal{D}\backslash\mathcal{D}_\text{feas}$ denotes the set of generated designs that cannot fulfill the target in reality. In an ideal scenario, all solutions in $\mathcal{D}$ would satisfy the target, which means $\mathcal{D} = \mathcal{D}_\text{feas}$, and their density profiles should coincide. However, this ideal scenario is impossible in reality due to both limited model accuracy and uncertainty. As RIGID generates designs by sampling the entire design space based on the likelihood values, it is expected for non-satisfactory designs to be generated as long as the likelihood values for these designs are non-zeros. Therefore, the fact that a design is generated does not mean that the model is certain about the design satisfying the target. We have to look at the estimated likelihood to know how likely the target will be satisfied and make decisions from there. RIGID thus incorporated the model's confidence into the inverse design process.
For a reasonable model, most designs in $\mathcal{D}\backslash\mathcal{D}_\text{feas}$ should have low estimated likelihood values. Consequently, the density of $\mathcal{D}_\text{feas}$'s likelihood is a result of shifting some of $\mathcal{D}$'s density from left (low likelihood) to right (high likelihood). This expectation aligns with the observation in Fig.~\ref{fig:acoustic}D.

When sampling new designs or selecting solutions from generated designs, we can put a \textit{sampling threshold} $\tau \in (0, 1)$ on the likelihood values to filter out ``less promising'' solutions. To further examine model behavior and quantify how $\tau$ affects the inverse design outcome, we define the following metrics:
\begin{equation}
\begin{split}
\text{Selection Rate} & = \frac{|\mathcal{D}_{\phi \geq \tau}|}{|\mathcal{D}|}, \\
\text{Satisfaction Rate} & = \frac{|\mathcal{D}_{\phi \geq \tau} \cap \mathcal{D}_\text{feas}|}{|\mathcal{D}_{\phi \geq \tau}|}, \\
\text{Average Score} & = \frac{1}{|\mathcal{D}_{\phi \geq \tau}|} \sum_{i=1}^{|\mathcal{D}_{\phi \geq \tau}|} q_i, \\
\end{split}
\label{eq:metrics}
\end{equation}
where $\mathcal{D}_{\phi \geq \tau}$ is the set of generated designs with the likelihood of at least $\tau$ (i.e., the selected designs) and $q_i$ denotes the percentage overlap between the target and the actual behavior of selected designs.
The satisfaction rate evaluates how many selected designs satisfy the target based on a binary criterion (i.e., whether or not a design satisfies the complete target), whereas the average score provides a soft measure where partial satisfaction is also counted. The average score is lower-bounded by the satisfaction rate.

As shown in Fig.~\ref{fig:acoustic}E, the selection rate decreases when $\tau$ increases since more solutions are filtered out. On the other hand, both the satisfaction rate and the average score increase with $\tau$, which indicates a high correlation between the estimated likelihood of a solution and its probability of actually achieving the target. As $\tau$ reaches 0.6, the satisfaction rate and the average score reach 1, indicating that all generated designs satisfy their targets. When sampling or selecting new solutions, we can use the sample threshold $\tau$ to tune the trade-off between exploitation and exploration~\textemdash~a low $\tau$ favors exploration as sampled solutions will cover a larger area of the design space, while a high $\tau$ favors exploitation as sampled solutions will have a higher chance of satisfying the target. 

\begin{figure}[h]
\centering
\includegraphics[width=1\textwidth]{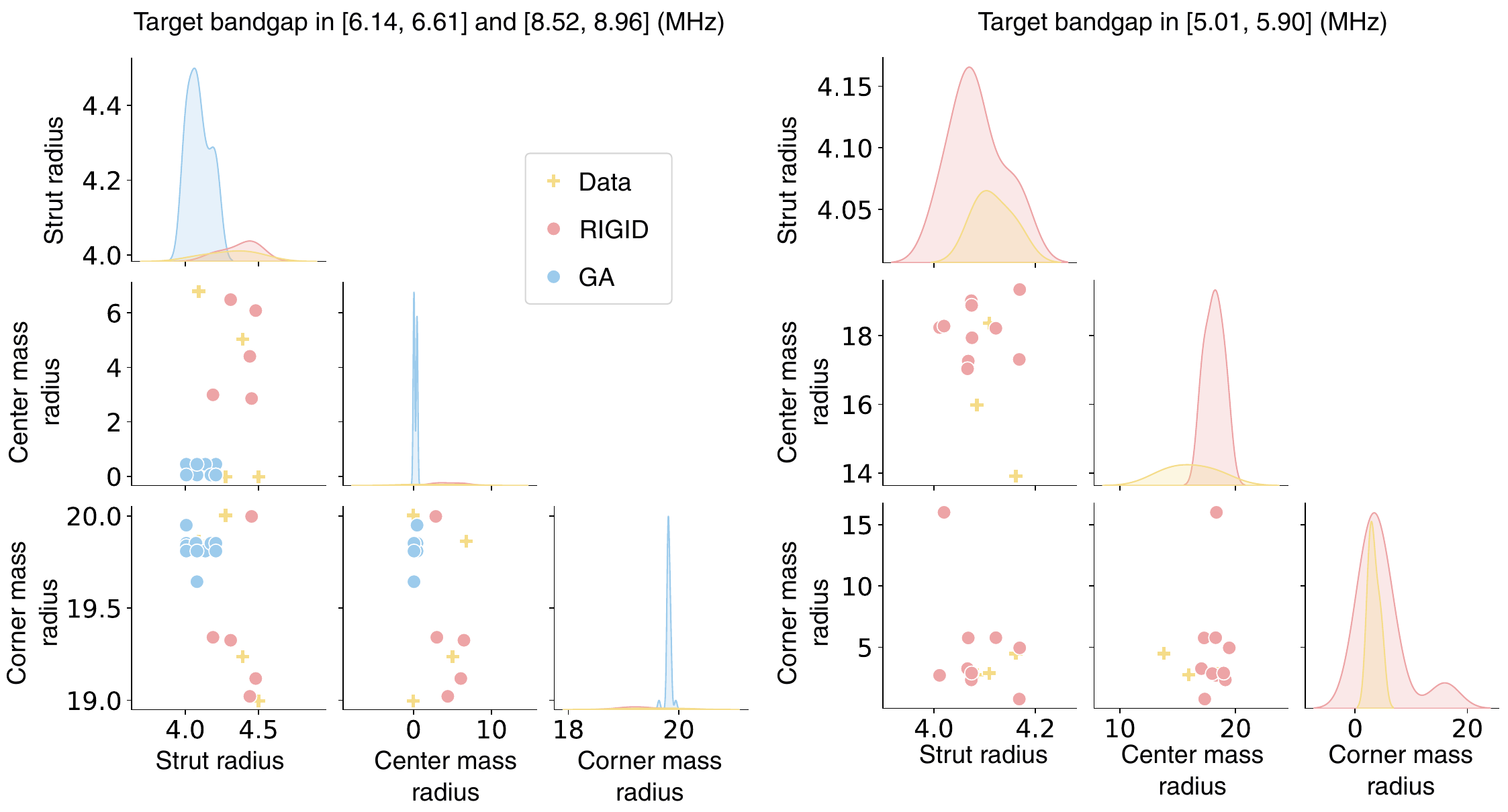}
\caption{\revised{Distributions of satisfactory solutions for two bandgap targets. The off-diagonal plots show the pairwise bivariate distributions of design variables, and the diagonal plots show the marginal distributions of the data in each column. The left panel shows that GA designs are highly localized while RIGID can lead to diverse solutions. The right panel indicates that none of the GA designs satisfy the target, while satisfactory RIGID designs are diverse and can be very different from feasible designs from data. Solutions from data include feasible designs in both training and test data.}
}
\label{fig:acoustic_pairplots}
\end{figure}

\revised{In the comparative study, we generated the same number of designs for the same targets by applying GA. We computed the satisfaction rate of GA designs and found that the value is close to the satisfaction rate of RIGID designs when setting $\tau$ to 0.2. This means that if we set $\tau$ to be above 0.2, RIGID will most likely achieve a higher satisfaction rate than GA. The distributions of generated satisfactory solutions shown in Fig.~\ref{fig:acoustic_pairplots} indicate that GA designs are highly localized while RIGID can capture the high diversity of the inverse design solutions, and sometimes discover solutions very different from the original feasible solutions from data. Besides, Fig.~\ref{fig:acoustic_pairplots} also shows an example where GA fails to find any satisfactory solution while RIGID can still generate a very diverse set of satisfactory solutions.
}

Figure~\ref{fig:acoustic}F visualizes the geometries and dispersion relations of RIGID designs generated based on a randomly created bandgap target. Only the top five designs with the highest likelihood values are shown.
In this example, our method generates geometrically different designs that have a high probability of achieving target bandgaps, each yielding a slightly different dispersion relation. This is promising in design applications requiring other material properties, such as dynamic wave velocity or quasi-static stiffness, in which the user can select from a menu of designs with the same target bandgap but other varying properties. Generated designs based on the other nine bandgap targets can be found in \textit{SI Appendix, Figs.~S2-S6}.

\subsection{Applying RIGID to Design Optical Metasurfaces with Target High-Absorbance Wavelengths}
\label{sec:optical}

Optical metasurfaces are artificially engineered systems that can support exotic light propagation building on subwavelength inclusions~\cite{pendry2000negative, yu2011light, kildishev2013planar, cui2014coding, bukhari2019metasurfaces, hu2021review}. Among a diverse array of devices, metamaterial absorbers~\cite{landy2008perfect, liu2010infrared, hao2010high, watts2012metamaterial, cui2014plasmonic, liu2017experimental, lee2018complete} have been intensely studied for medical imaging, sensing, and wireless communications.

\begin{figure}
\centering
\includegraphics[width=0.8\textwidth]{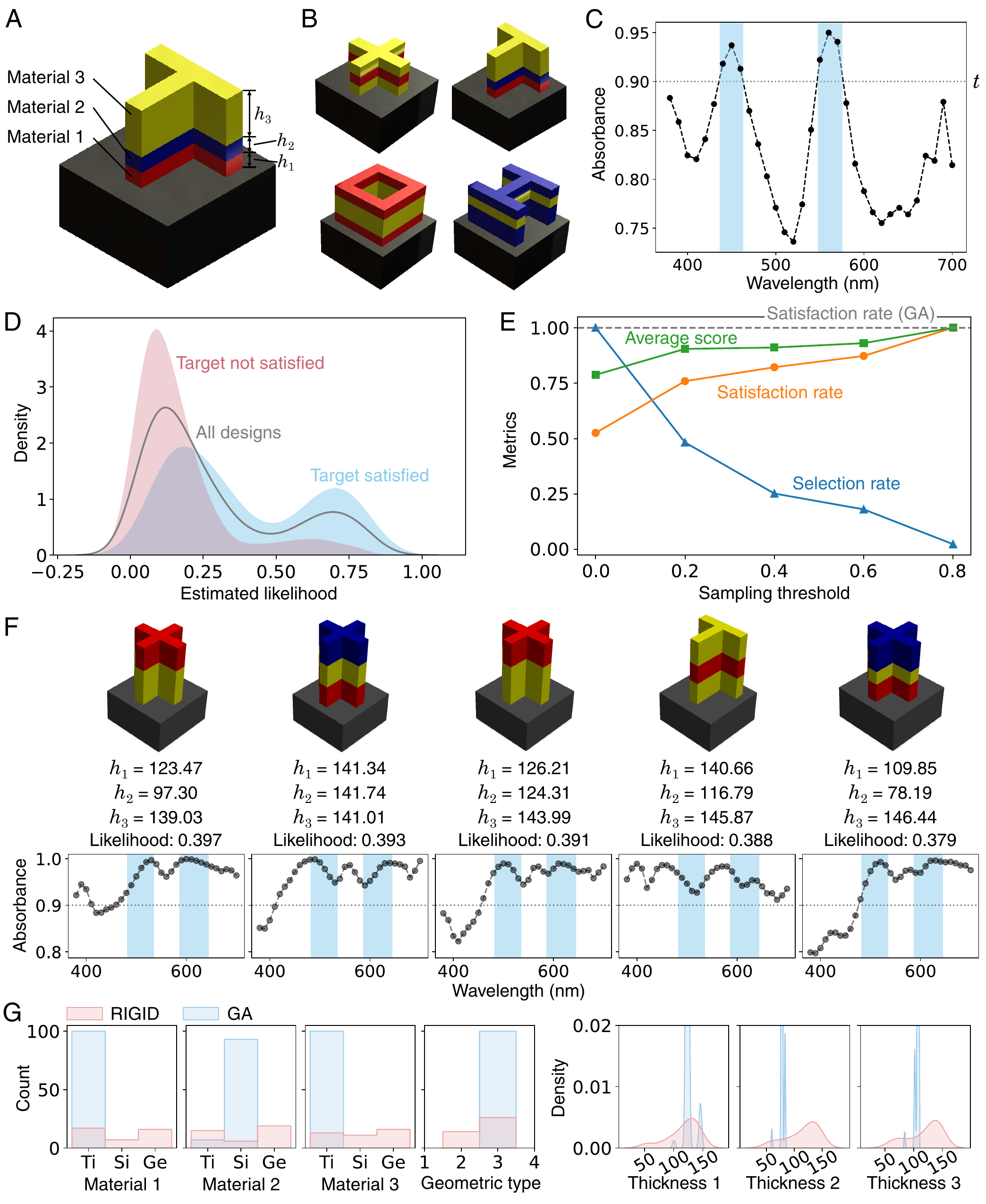}
\caption{Optical metasurface design problem configuration and results. (A-B)~Design variables (materials, layer thicknesses, and cross-section geometry types). (C)~A sample absorbance spectrum and the wavelength intervals (highlighted wavelength regions) corresponding to absorbance above the threshold $t$. The design objective is to generate new optical metasurface designs that exhibit higher absorbance than a threshold $t$ at the user-defined wavelength interval(s). (D)~KDE of the estimated likelihood for generated designs. \revised{(E)~Satisfaction rates, average scores, and selection rates for RIGID designs under varying sampling thresholds (solid lines), in comparison to the satisfaction rate of GA designs (horizontal dashed line).} (F)~Designs (geometries and material selections) and corresponding absorbance spectra of five metasurfaces generated by RIGID. These five solutions are generated designs with the highest likelihood of satisfying specified target high-absorbance regions (marked as highlighted wavelength regions). All the layer thicknesses ($h_l, l=1, 2, 3$) have a unit of nm. Here all five designs satisfy the target. Generated designs for other targets are shown in \textit{SI Appendix, Figs.~S7-S9}. \revised{(G)~Distributions of design variables for satisfactory solutions generated by RIGID and GA (for the same target defined in Panel F). GA designs are highly localized and lack diversity compared to RIGID designs.}
}
\label{fig:optical}
\end{figure}

In this case study, we consider four types of cross-sections ($c \in \{1,2,3,4\}$) chosen from the literature (Fig.~\ref{fig:optical}B). It is assumed that a 3D geometric instance is composed of a stack of three layers of prismatic unit cells, each of which is vertically extruded and stacked (Fig.~\ref{fig:optical}A). The geometries constructed in this way can be regarded as an instantiation of multilayered metasurfaces~\cite{zhou2018multilayer, li2018wideband, marino2021harmonic, malek2022multifunctional, zhang2023high}, which offer richer design freedom than the single-layer counterpart. The thickness of each layer ($h_l, l=1, 2, 3$) is allowed to continuously vary between 50 and 150 nm. Herein we do not consider parametric variations of a given type of unit cell cross-section; yet those can be trivially incorporated in the proposed design framework if necessary.

We also design the material of each layer ($m_l, l=1, 2, 3$). Three dielectric materials of interest, each of which is assigned to a different color in Fig.~\ref{fig:optical}A, are Ti (red), Si (blue), and Ge (yellow). 
The challenge associated with this design problem is its mixed-variable design space containing three continuous variables (i.e., layer thicknesses) and four categorical variables (i.e., material choices at each layer and the geometry type), which may lead to potentially complicated underlying design-response relations (e.g., those across different geometry or material types).
In general, a dielectric material is characterized through a complex refractive index $ \Tilde{n} \in \mathbb{C}$ defined as $\Tilde{n} = n + j k$, where $j=\sqrt{-1}$ is the imaginary unit, $n\in\mathbb{R}$ involves the speed at which the light propagates through the material, and $k\in\mathbb{R}$ is the extinction coefficient that dictates the energy loss due to the material.
Within the frequency regime of interest, those exhibit nonlinear dispersion; both the real and imaginary terms in general are a non-analytic function of excitation wavelength $s$, i.e., $n(s)$ and $k(s)$. In addition, the impact of the same material choice on the spectral response $A(s)$ varies depending on the layer location at which the material is placed. 

Based on the above configuration, we randomly sampled 258 sets of design variables $\mathbf{x} = (c, h_1, h_2, h_3, m_1, m_2, m_3)$ and computed their corresponding absorbance spectra using wave analysis. We set $t=0.9$ as the absorbance threshold, so that ``high'' absorbance means the absorbance $A(s)$ is no less than 0.9. We trained a random forest on 206 training data (i.e., 80\% of the 258 designs and corresponding absorbance spectra) to predict whether ``high'' absorbance is presented (i.e., the binary indicator $y=1$) at a wavelength $s$ for a given design $\mathbf{x}$. The trained random forest has a test F1 score of 0.83. The confusion matrix on test data is shown in \textit{SI Appendix, Tab.~S2}.

Note that this problem involves inverse design with both continuous and categorical variables, which common optimization and generative modeling-based inverse design cannot handle well without special treatment~\cite{zhang2020bayesian,ma2020vaem,xu2018synthesizing}. On the other hand, our random forest-based method can naturally address such mixed-variable problems without any issues.

Similar to the acoustic metamaterial design problem, we use 10 randomly created targets to evaluate the inverse design performance of RIGID, except that here a target is represented as the wavelength range(s) within which absorbance should be at least 0.9. We generated 100 designs for each target by sampling from the estimated likelihood distribution. Among the 1,000 generated solutions, we successfully conducted wave analysis for 911 designs and obtained their absorbance spectra. Figure~\ref{fig:optical}D shows the KDE for the likelihood of these 911 designs, conditioned on their target satisfaction. The densities share similar behavior as in the acoustic problem (Fig.~\ref{fig:acoustic}D)~\textemdash unsatisfied/infeasible designs $\mathcal{D}\backslash\mathcal{D}_\text{feas}$ are concentrated at low likelihood regions, which causes the likelihood density of satisfied/feasible designs $\mathcal{D}_\text{feas}$ to be a result of shifting some of $\mathcal{D}$'s density from left (low likelihood) to right (high likelihood). The sampling threshold and metrics relation shown in Fig.~\ref{fig:optical}E also follow the same trend as in the acoustic problem (Fig.~\ref{fig:acoustic}E), which again demonstrates a strong positive correlation between the estimated likelihood and the probability of generated designs actually achieving their targets.

\revised{We also generated the same amount of designs under the same targets by using GA. We found that GA designs can generate 100\% satisfaction rate (Fig.~\ref{fig:optical}E). 
However, similar to the acoustic metamaterial design example, the generated designs are localized compared to RIGID designs which cover much wider feasible region(s) in the design space. This can be reflected by the design variable distributions of generated feasible solutions shown in Fig.~\ref{fig:optical}G, which corresponds to generated designs for a randomly created target (the target is shown as the highlighted wavelength regions in Fig.~\ref{fig:optical}F). For example, all the GA designs have the same cross-section geometric type while RIGID generated satisfactory designs with different geometric types.
It is easy to see that, without considering solution diversity, achieving a high target satisfaction rate is trivial\textemdash as long as we discover one satisfactory solution, we can add sufficiently-small perturbations to that solution to generate an infinite number of alternative solutions that have similar responses and thus are also likely to satisfy the target.
Note that compared to the acoustic problem, this optical problem is less challenging as there is an average of 28.5\% data already satisfying the ten targets (in contrast to an average of 9.3\% satisfactory designs from data in the acoustic problem). Therefore, finding a few satisfactory solutions for the optical problem is relatively easy; the challenge lies in discovering all feasible solutions, where RIGID excels by identifying a wide range of them.
We can also see that when the sampling threshold is 0.8, RIGID achieves a satisfaction rate of 100\%, demonstrating its ability to trade off the diversity of generated designs for higher feasibility.
}

Figure~\ref{fig:optical}F shows generated RIGID designs with the top five likelihood estimations for a randomly created target. While the materials, cross-section geometries, and layer thicknesses of generated designs can be different, all the designs satisfy the target. 
Generated designs based on the other nine targets can be found in \textit{SI Appendix, Figs.~S7-S9}.

\subsection{Synthetic Design Problems for Rapid Validation and Visualization}
\label{sec:synthetic}

While the above metamaterial design problems represent practical use cases, the validation study is time-consuming due to the expensive computation of metamaterials' responses such as dispersion relations and absorbance spectra. To allow fast validation of the proposed method and easier inspection of the estimated likelihood in the design space, we create two synthetic case studies. Both problems have 2-dimensional ``design spaces'' that allow easy visualization.

\begin{figure}
\centering
\includegraphics[width=0.8\textwidth]{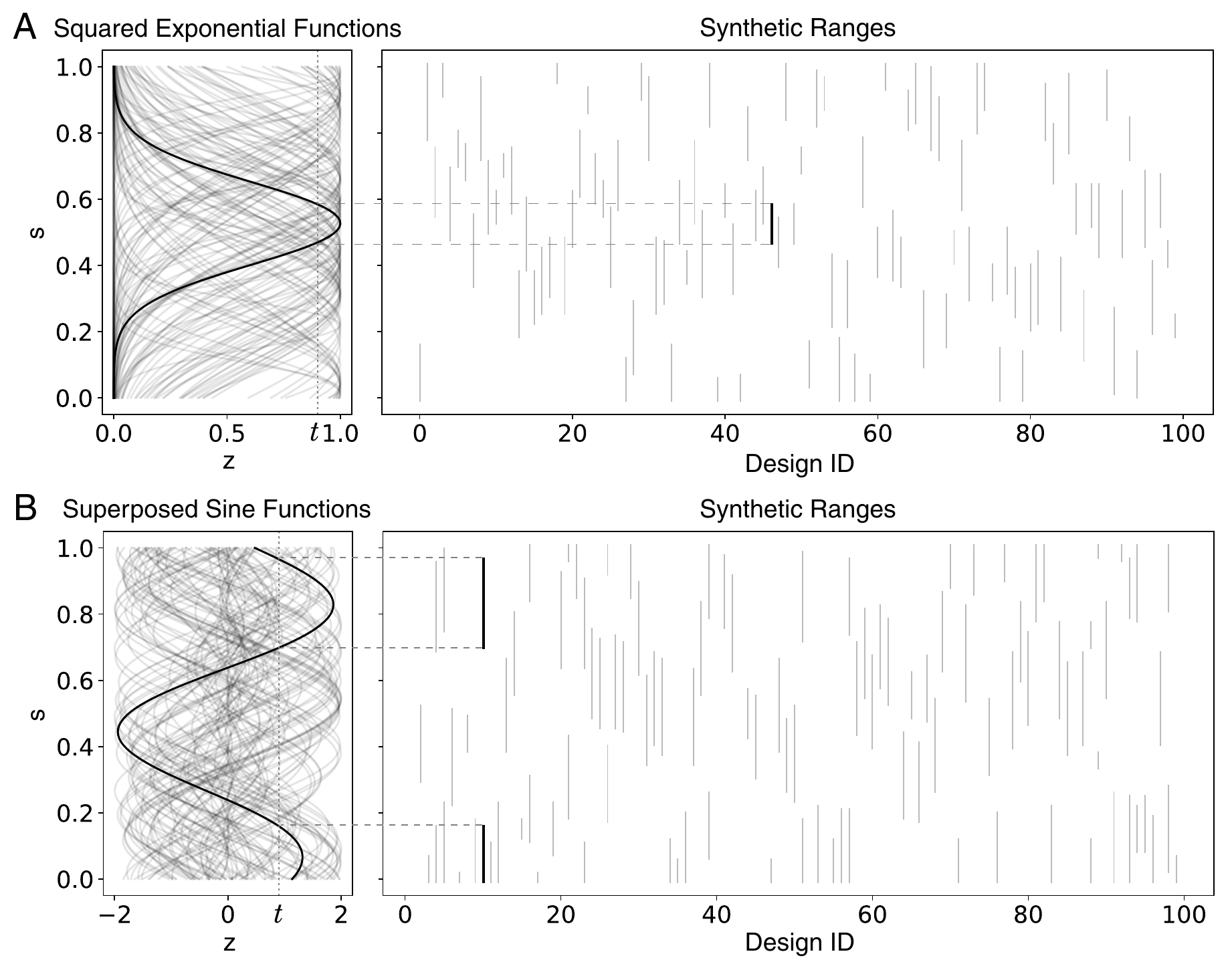}
\caption{Synthetic data creation for (A) the SqExp problem and (B) the SupSin problem. For each problem, the left panel shows 100 functions with randomly sampled parameters $a$ and $b$. We treat $a$ and $b$ as synthetic design variables, and the corresponding functions as quantitative responses (e.g., absorbance spectra of optical metasurfaces). The right panel shows qualitative responses (e.g., high-absorbance wavelength ranges or bandgaps) are simulated by synthetic ranges, derived by thresholding the 100 functions (Equations~\ref{eq:test_func_sqexp} and \ref{eq:test_func_sin} with threshold $t$=0.9).}
\label{fig:synthetic_data}
\end{figure}

\paragraph{SqExp Problem} To construct the first synthetic problem, we used a \textit{squared exponential function} with tunable parameters $a$ and $b$ to mimic the quantitative functional response of metamaterials. The qualitative response (e.g., ``high'' or ``low'' energy absorption at a wavelength) is defined as: 
\begin{equation}
I(a,b; s) =
    \begin{cases}
      0, & \text{if $z = \exp\left(-{\left(\frac{s-a}{0.3b+0.1}\right)^2}\right) < t$} \\
      1, & \text{otherwise}
    \end{cases},
\label{eq:test_func_sqexp}
\end{equation}
where $z$ represents quantitative response and $t$ is a threshold that converts $z$ into a qualitative response $I(a,b; s)$. Specifically, $I(a,b; s) = 1$ can mean the existence of a bandgap or high absorbance at a frequency $s$. Then $\{s|I(a,b; s) = 1\}$ represents a range of $s$ that mimics our interested material behavior, such as the frequency range of the bandgap or the wavelength range of high absorbance. We call this the \textit{positive range}. By varying $a$ and $b$, we can produce different synthetic responses and positive ranges. Therefore, we can use $a$ and $b$ as synthetic design variables. There is a clear relation between these design variables and the positive range that Eq.~\ref{eq:test_func_sqexp} creates~\textemdash~$a$ and $b$ control the center location and the width of the positive range, respectively.

In this design problem, we sampled 100 sets of $a$ and $b$ uniformly at random. We set $t$ as 0.9. Based on Eq.~\ref{eq:test_func_sqexp}, we obtained the corresponding responses (Fig.~\ref{fig:synthetic_data}A). These sets of $a$, $b$, and responses constitute a dataset for training and testing our model. 

\paragraph{SupSin Problem} Another synthetic design problem was constructed by replacing
the squared exponential function in the SqExp problem with a \textit{superposed sine function}. Given synthetic design variables $a$ and $b$, we can produce qualitative responses using the following equation:
\begin{equation}
I(a,b; s) =
    \begin{cases}
      0, & \text{if $z = \sin\left(2\pi(s+a)\right) + \sin\left(3\pi(s+b)\right) < t$} \\
      1, & \text{otherwise}
    \end{cases}.
\label{eq:test_func_sin}
\end{equation}
Same as in the SqExp problem, we set $t=0.9$ and created a dataset with 100 sets of synthetic design variables and corresponding ranges derived from synthetic responses (Fig.~\ref{fig:synthetic_data}B). 
Unlike the squared exponential function, the superposed sine function can be multimodal, which means it can result in multiple synthetic ranges to mimic, for example, multiple bandgaps. The bandgap locations are controlled by $a$ and $b$.

\begin{figure}[h]
\centering
\includegraphics[width=1\textwidth]{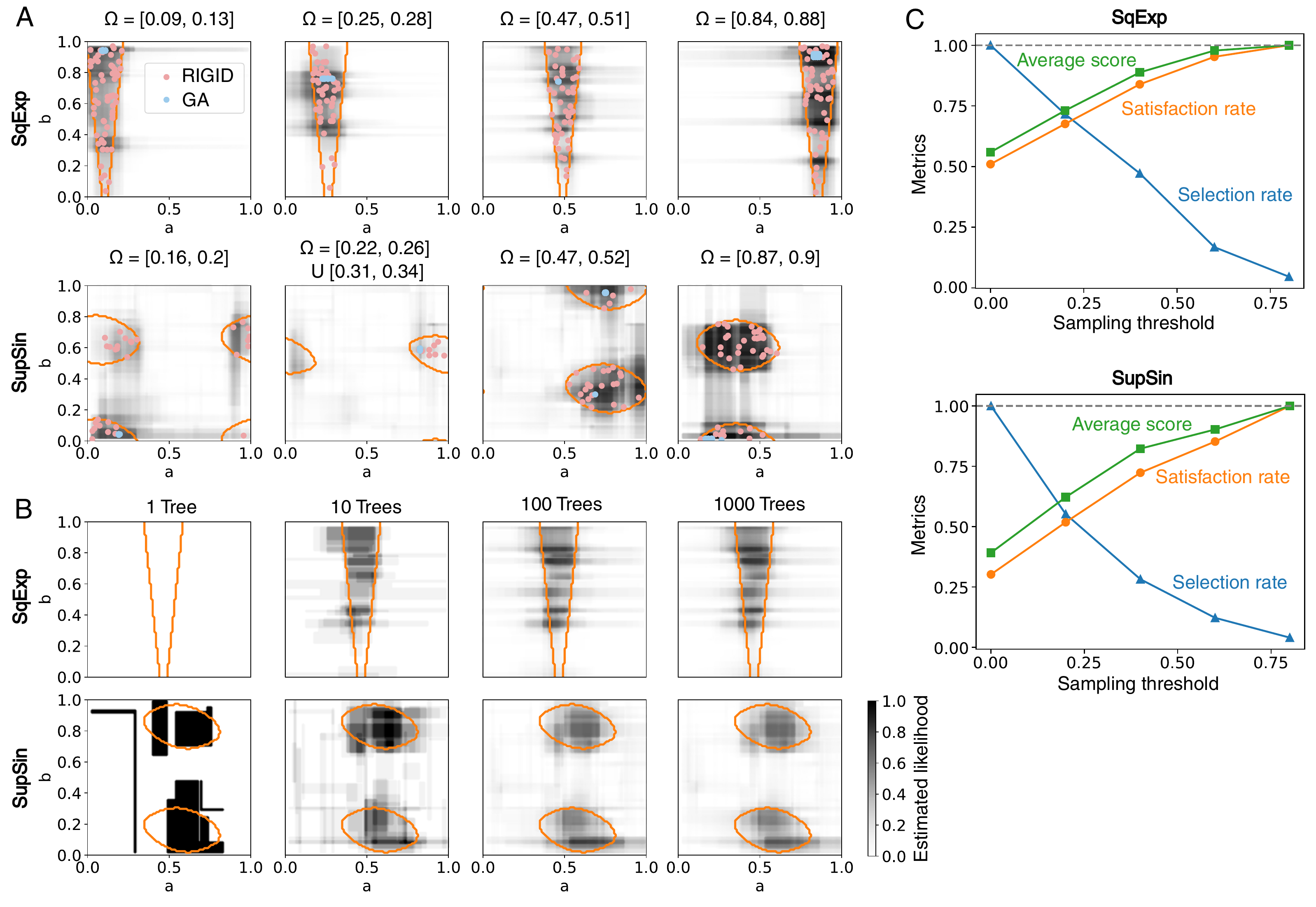}
\caption{Visualization of estimated likelihood and validation metrics for synthetic problems. (A)~Likelihood function values for randomly created design targets. Orange lines show boundaries of actual feasible regions associated with the targets $\mathcal{T} = \{I(a, b; s) = 1 | \forall s \in \Omega\}$. \revised{Points show satisfactory RIGID designs and GA designs.} (B)~Likelihood function values estimated by random forests with varying numbers of decision trees. The design target is set as $\mathcal{T} = \{I(a, b; s) = 1 | \forall s \in [0.45, 0.48]\}$ for the SqExp problem and $\mathcal{T} = \{I(a, b; s) = 1 | \forall s \in [0.63, 0.68] \cup [0.69, 0.71]\}$ for the SupSin problem. (C)~Validation metrics for inverse design generation.}
\label{fig:synthetic_results}
\end{figure}

For each synthetic example, we split the data into 80 training data and 20 test data. We trained a random forest with the same hyperparameter settings as the other problems, to predict the binary indicators $I(a,b; s)$. The F1 scores are 0.85 and 0.86 for the SqExp and the SupSin problems, respectively. The resulting confusion matrices are shown in \textit{SI Appendix, Tables~S3-S4}. We evaluated the inverse design performance with the trained models.

Due to the fast evaluation of Equations~\ref{eq:test_func_sqexp} and \ref{eq:test_func_sin}, we can exhaust all the possible solutions in the design space to obtain the ground-truth feasible region(s) for a target. Figure~\ref{fig:synthetic_results}A shows the estimated likelihood values and the ground-truth feasible regions under randomly created targets. In general, high-likelihood regions match actual feasible regions well, which further demonstrates the effectiveness of RIGID. We can also observe that feasible regions in the SqExp and the SupSin problems follow distinct patterns. In the SqExp problem, $a$ and $b$ control the center location and the width of the positive range, respectively. Therefore, the position of the feasible region along the $a$-axis moves with the location of the target range, while the feasible region gradually shrinks as $b$ decreases since the decrease of $b$ (i.e., positive range width) causes the choice of $a$ (i.e., positive range center location) to be more restricted to fit the target range. In the SupSin problem, there might be multiple positive ranges appearing at the peaks of the superposed sine function in Eq.~\ref{eq:test_func_sin}. Design variables $a$ and $b$ control positive range locations by translating each sine function. Due to the sine function's periodicity, we can obtain multiple feasible regions along both $a$- and $b$-axes.
Figure~\ref{fig:synthetic_results}A shows that the likelihood estimation by RIGID successfully captured the above-mentioned patterns of feasible regions.

\revised{We also visualize the distributions of satisfactory designs generated by both RIGID and GA in Fig.~\ref{fig:synthetic_results}A. Again, we observe a much broader coverage of the feasible design space by RIGID designs, while GA designs are merely minor perturbations of one or two feasible solutions.}

Figure~\ref{fig:synthetic_results}B demonstrates how the estimated likelihood varies when increasing the number of trees in a random forest. With a single decision tree, the estimated likelihood function is almost a binary function and highly inaccurate. The likelihood in the SqExp case is even zero everywhere, which makes it impossible to sample designs based on the likelihood. As the number of trees increases, the likelihood function becomes smoother and eventually converges.

Besides these qualitative visual inspections, we also calculated the metrics proposed in Eq.~\ref{eq:metrics}, as shown in Fig.~\ref{fig:synthetic_results}C. For each of the two synthetic problems, these metrics were computed on 500 designs generated by giving five random target ranges. Again, the satisfaction rate and the average score increase with the sampling threshold, indicating a strong correlation between the sampling threshold and the probability of generated designs actually achieving their targets. In both problems, all the selected designs satisfy their targets (i.e., the satisfaction rates and average scores reach 1) when the sampling threshold reaches 0.8.

\section{\revised{Discussion}}

\revised{In this section, we discuss the benefits and necessities of RIGID in terms of its interpretability, data demand, the ability to deal with discrete design variables, and the diversity of generated designs.}

\subsection{\revised{Interpretability}}

\revised{RIGID offers two levels of interpretability. Firstly, RIGID is more transparent than deep learning models. The transparency of the random forest as a forward model helps us understand how the predictions are made. The inverse design process based on likelihood estimation and sampling is also transparent. The transparency of both forward and inverse processes offers insights into the decision-making process and diagnostic capabilities. Secondly, owing to the interpretability of the forward prediction model, we can derive the likelihood, which the inverse design relies on, by probing into each decision path of trees in the random forest. The likelihood is an interpretable metric to help us understand the probability of a generated design satisfying the given target and thus aid the decision-making process when selecting final solutions. The results (Figures~\ref{fig:acoustic}D, \ref{fig:acoustic}E, \ref{fig:optical}D, \ref{fig:optical}E, and \ref{fig:synthetic_results}C) have demonstrated a high correlation between the likelihood and actual probability of target satisfaction. 
The likelihood also serves as an indicator of the need for adjusting the design space. When the likelihood is low across the entire design space, the model believes the target is almost impossible to achieve, which may indicate a need for expanding the design space (i.e., extending design space bounds or adding new design freedom) to allow wider coverage of the response space and easier discovery of satisfying solutions. Please see more discussion in \textit{SI Appendix, Sec.~S4}. Another benefit of having the explicit likelihood as an interpretable metric is that we can adjust the likelihood threshold to trade-off between exploration and exploitation~\textemdash~higher (lower) threshold means more focus on exploitation (exploration). Both levels of interpretability were absent from most prior works of iteration-free, single-shot inverse design methods.}

\subsection{\revised{Data Demand}}

\revised{RIGID is particularly useful when data collection is expensive, as in many cases where high-fidelity simulation or experimental data are needed. 
Despite the advancement of surrogate solvers and hardware-accelerated simulation, a persistent trade-off between computational resources and accuracy remains, exacerbated by challenges in accessing high-performance infrastructure and collecting high-fidelity data.
With most of the prior inverse design works based on deep learning with a high demand for data and computational resources, it is useful to create a method that works under scenarios at the other end of the data/computational resource requirement spectrum. Additionally, 
when the only way to acquire high-fidelity data is through experiments, data deficiency will become a pressing challenge for deep learning-based methods.
While there is a large collection of deep learning methods to address single-shot inverse design problems with large datasets, the setting of small data seems to be under-studied by prior works, despite this setting's practicality under constrained budgets. RIGID fills this gap by offering a much more data-efficient and interpretable alternative to deep-learning-based single-shot inverse design methods.}

\subsection{\revised{Diversity versus Feasibility of Generated Designs}}

\revised{The MCMC in RIGID can sample new designs in the entire design space based on the estimated likelihood, which is not limited to local optimality and has the potential to discover multiple very different satisfying solutions (if any) in one shot. This is clearly demonstrated in the SupSin example where there are multiple disconnected feasible regions in the design space and the likelihood estimated by RIGID successfully coincides with these feasible regions (Fig.~\ref{fig:synthetic_results}A). The ability to generate multiple satisfying solutions allows us to consider other figures of merit, such as manufacturability, compatibility, and cost, in addition to the main objective of satisfying the target response. Besides enabling flexibility of design choice, generating a comprehensive solution set can also bring more insights into the structure-property relation associated with the target. For example, after obtaining a diverse set of designs satisfying the same target, we can investigate the differences among these designs to identify critical features that lead to the satisfaction of the target.}

\revised{
For all four problems demonstrated in this work, when the sampling threshold is set to 0.8, RIGID achieves a satisfaction rate of 100\%.
Although GA can also lead to a high target satisfaction rate for relatively simple inverse design problems (e.g., the optical metasurface design problem in Sec.~\ref{sec:optical}), the generated solutions lack diversity. In contrast, RIGID provides a unique way to control the trade-off between the diversity and the feasibility of generated designs by tuning the sampling threshold.
It is important to note that, without considering diversity, a high target satisfaction rate becomes trivial, as we can simply apply sufficiently-small perturbations to one feasible solution to generate an infinite number of similar solutions that have similar responses and hence are also likely to meet the target. In many inverse design problems where solutions are non-unique, finding one solution can be simple, but the real challenge lies in discovering feasible solutions that cover a broad range of the design space\textemdash a problem where RIGID will outperform optimization-based methods as demonstrated by our results.
}

\subsection{\revised{Discrete Design Variables}}

\revised{The use of a random forest in RIGID allows easy consideration of discrete design variables. For example, our optical metasurface design problem has the geometry type and the material choice for each layer as categorical design variables. In contrast, these discrete variables will pose challenges for other inverse design methods based on some machine learning models (e.g., conditional generative models) or gradient-based optimization. 
}

\section{Summary and Outlook}
\label{sec:conclusion}

We proposed RIGID, an iteration-free, single-shot inverse design method that generates metamaterials to satisfy qualitative behaviors of functional responses. Such qualitative behaviors are important design targets in many applications such as tailoring bandgaps of acoustic metamaterials for wave-guiding, focusing, and vibration mitigation, or tailoring the absorption level of optical metasurfaces at certain wavelengths for medical sensing, imaging, and communication applications. Unlike most existing machine learning-based inverse design methods that require training an inverse model to map targets to designs, the RIGID method takes advantage of the random forest's interpretability and derives the likelihood of target satisfaction by probing the trained forward model. Incorporated with MCMC, one can sample a desired number of new designs based on the estimated likelihood. Therefore, RIGID functions as a generative model that can capture the conditional distribution of satisfying designs given a target, or in other words, the one-to-many mapping from the target to satisfying designs. Using both real-world and synthetic design problems, we demonstrated that RIGID is efficient and effective on datasets with training sample sizes smaller than 250. We used both qualitative and quantitative approaches to validate the proposed method. The quantitative results revealed a strong correlation between the estimated likelihood of a solution and its probability of actually achieving the target, which demonstrated the effectiveness of the likelihood estimation. Due to the fast evaluation of output responses and the transparency of ground-truth solutions, the proposed synthetic problems can be used for future benchmarking studies of metamaterial design problems.

While we address qualitative design targets in this study, a similar idea has the potential to generalize to quantitative targets. Such problems can be, for example, generating optical metasurface designs with specific optical spectra~\cite{liu2018generative, ma2019probabilistic}, generating functional materials with target nonlinear constitutive relations~\cite{lew2023single,vlassis2023denoising}, or generating programmable metamaterials with prescribed functional responses~\cite{li2022digital,lin2023mechanical}. It is also straightforward to adjust the target to achieve multifunctionality (e.g., negative/positive Poisson’s ratio under low/high compression rate~\cite{bossart2021oligomodal}).

Although this study only demonstrates the RIGID method on parametric design (i.e., designs are represented by geometric and/or material parameters), the method also applies to shape or topological design problems where the shape or topology of designs can vary without being restricted to a limited number of geometric parameters~\cite{liu2018generative,jiang2019free,ma2019probabilistic,wen2020robust,an2021multifunctional}. In those cases, as valid designs only lie on a lower-dimensional manifold of the design space, the likelihood of target satisfaction will be zero almost everywhere in the original design space. Thus before applying RIGID, we need to obtain a latent representation that compactly captures the manifold of valid designs~\cite{wang2020deep,chen2023gan}, and use the latent representation as design variables for inverse design.

\section{Experimental Section}
\label{sec:experimental_secion}

\paragraph{\revised{Configurations for GA}} \revised{The initial population for GA is generated by randomly sampling solutions within the design variable bounds. At each generation, we select individuals to reproduce for a new generation by randomly picking three individuals from the population, selecting the best one, and repeating this process 300 times. We set the probability for crossover and mutation to be 0.5 and 0.2, respectively. Once an individual is mutating, there is a 5\% chance for each design variable in this individual to mutate. For a continuous design variable, we define the mutation operation as adding Gaussian noise with a mean of 0 and a standard deviation equal to 1\% of the variable's range. For a categorical design variable, mutation is defined as randomly reassigning the category of the variable. We ran GA for 10 generations, each with a population of 300. Finally, we select a set of solutions with the highest objective values among all the individuals created during GA, and use these selected solutions as generated designs.
}

\paragraph{Computation of Acoustic Dispersion Relations} We performed Bloch-wave analysis in COMSOL Multiphysics to compute the dispersion relations of acoustic metamaterials. Poisson's ratio of 0.49, Young's modulus of 2.7 GPa, and density of 1170 kg/m${^3}$ were set as material properties with $\sim1.5\times10^4$ mesh elements per unit cell. 
We used Floquet-Bloch periodic boundary conditions to obtain the first 60 eigenfrequencies along all symmetry domains of the cubic irreducible Brillouin zone (Fig. \ref{fig:acoustic}B) for all lattices, thus generating a dispersion relation. 

\paragraph{Computation of Optical Absorbance Spectra} We computed the absorbance spectra for optical metasurfaces using wave analysis inspired by Zhang et al.~\cite{zhang2023high}. The RF Module of COMSOL Multiphysics\textsuperscript{\textregistered}~\cite{comsol2020} was used to evaluate the spectral response of concern, which is the energy absorbance $A(s)$ in the visible regime (380-700 nm). An absorbance spectrum is computed with respect to 33 wavelength components $s_k$ that are uniformly discretized over the specified range. An incident plane wave is assumed to be given from the port, located at the top face of the analysis domain. We set the periodicity of the analysis domain as 400 nm. The periodic boundary condition on electromagnetic fields is imposed on the lateral faces of the analysis domain. A substrate made of SiO$_2$ is placed right below a given unit cell instance (the black layers in Fig.~\ref{fig:optical}A-B). With full electric fields computed through the wave analysis, the energy absorbance at a single wavelength $s$ is quantified as $A(s)=1-|S_{11}(s)|^2$, where $S_{ij}$ is the component of the $S$-parameter matrix that specifies energy transfer between ports. We use the data presented in Ref.~\cite{aspnes1983dielectric} to set the material dispersion of the dielectric.

\section*{Acknowledgement}

This work was supported by the startup funds from the J. Mike Walker '66 Department of Mechanical Engineering at Texas A\&M University, the National Science Foundation (NSF) BRITE Fellow program (CMMI 2227641), the NSF CSSI program (OAC 1835782), the Kansas City National Security Campus (PDRD \#705288), and NSF CAREER Award (CMMI-2142460). R.S. acknowledges financial support from the NSF Graduate Research Fellowship Program.

\section*{Data Availability Statement}

Source code and data for this work are available at \url{https://github.com/DIGITLab23/RIGID}.

\section*{Keywords}

metamaterials, inverse design, generative design, interpretable machine learning, functional response

\bibliographystyle{unsrt}
\bibliography{references}

\begin{thebibliography}{10}

\bibitem{kadic2013metamaterials}
Muamer Kadic, Tiemo B{\"u}ckmann, Robert Schittny, and Martin Wegener.
\newblock Metamaterials beyond electromagnetism.
\newblock {\em Reports on Progress in physics}, 76(12):126501, 2013.

\bibitem{bertoldi2017flexible}
Katia Bertoldi, Vincenzo Vitelli, Johan Christensen, and Martin Van~Hecke.
\newblock Flexible mechanical metamaterials.
\newblock {\em Nature Reviews Materials}, 2(11):1--11, 2017.

\bibitem{soukoulis2011past}
Costas~M Soukoulis and Martin Wegener.
\newblock Past achievements and future challenges in the development of
  three-dimensional photonic metamaterials.
\newblock {\em Nature photonics}, 5(9):523--530, 2011.

\bibitem{cummer2016controlling}
Steven~A Cummer, Johan Christensen, and Andrea Al{\`u}.
\newblock Controlling sound with acoustic metamaterials.
\newblock {\em Nature Reviews Materials}, 1(3):1--13, 2016.

\bibitem{schittny2013experiments}
Robert Schittny, Muamer Kadic, Sebastien Guenneau, and Martin Wegener.
\newblock Experiments on transformation thermodynamics: molding the flow of
  heat.
\newblock {\em Physical review letters}, 110(19):195901, 2013.

\bibitem{christensen2015vibrant}
Johan Christensen, Muamer Kadic, Oliver Kraft, and Martin Wegener.
\newblock Vibrant times for mechanical metamaterials.
\newblock {\em Mrs Communications}, 5(3):453--462, 2015.

\bibitem{pendry2000negative}
John~Brian Pendry.
\newblock Negative refraction makes a perfect lens.
\newblock {\em Physical review letters}, 85(18):3966, 2000.

\bibitem{gao2022acoustic}
Nansha Gao, Zhicheng Zhang, Jie Deng, Xinyu Guo, Baozhu Cheng, and Hong Hou.
\newblock Acoustic metamaterials for noise reduction: a review.
\newblock {\em Advanced Materials Technologies}, 7(6):2100698, 2022.

\bibitem{jiang2022flexible}
Shan Jiang, Xuejun Liu, Jianpeng Liu, Dong Ye, Yongqing Duan, Kan Li, Zhouping
  Yin, and YongAn Huang.
\newblock Flexible metamaterial electronics.
\newblock {\em Advanced Materials}, 34(52):2200070, 2022.

\bibitem{boley2019shape}
J~William Boley, Wim~M Van~Rees, Charles Lissandrello, Mark~N Horenstein,
  Ryan~L Truby, Arda Kotikian, Jennifer~A Lewis, and L~Mahadevan.
\newblock Shape-shifting structured lattices via multimaterial 4d printing.
\newblock {\em Proceedings of the National Academy of Sciences},
  116(42):20856--20862, 2019.

\bibitem{ma2022deep}
Chunping Ma, Yilong Chang, Shuai Wu, and Ruike~Renee Zhao.
\newblock Deep learning-accelerated designs of tunable magneto-mechanical
  metamaterials.
\newblock {\em ACS Applied Materials \& Interfaces}, 14(29):33892--33902, 2022.

\bibitem{casadei_piezoelectric_2012}
Filippo Casadei, Tommaso Delpero, Andrea Bergamini, Paolo Ermanni, and Massimo
  Ruzzene.
\newblock {Piezoelectric resonator arrays for tunable acoustic waveguides and
  metamaterials}.
\newblock {\em Journal of Applied Physics}, 112(6):064902, 09 2012.

\bibitem{liu2020acoustic}
Guang-Sheng Liu, Yang Zhou, Ming-Hao Liu, Ying Yuan, Xin-Ye Zou, and Jian-Chun
  Cheng.
\newblock Acoustic waveguide with virtual soft boundary based on metamaterials.
\newblock {\em Scientific Reports}, 10(1):981, 2020.

\bibitem{kim2021poroelastic}
Gunho Kim, Carlos~M Portela, Paolo Celli, Antonio Palermo, and Chiara Daraio.
\newblock Poroelastic microlattices for underwater wave focusing.
\newblock {\em Extreme Mechanics Letters}, 49:101499, 2021.

\bibitem{xie2018acoustic}
Yangbo Xie, Yangyang Fu, Zhetao Jia, Junfei Li, Chen Shen, Yadong Xu, Huanyang
  Chen, and Steven~A Cummer.
\newblock Acoustic imaging with metamaterial luneburg lenses.
\newblock {\em Scientific reports}, 8(1):16188, 2018.

\bibitem{matlack2016composite}
Kathryn~H Matlack, Anton Bauhofer, Sebastian Kr{\"o}del, Antonio Palermo, and
  Chiara Daraio.
\newblock Composite 3d-printed metastructures for low-frequency and broadband
  vibration absorption.
\newblock {\em Proceedings of the National Academy of Sciences},
  113(30):8386--8390, 2016.

\bibitem{krodel2015wide}
Sebastian Kr{\"o}del, Nicolas Thom{\'e}, and Chiara Daraio.
\newblock Wide band-gap seismic metastructures.
\newblock {\em Extreme Mechanics Letters}, 4:111--117, 2015.

\bibitem{bayat2018wave}
Alireza Bayat and Stavros Gaitanaros.
\newblock Wave directionality in three-dimensional periodic lattices.
\newblock {\em Journal of Applied Mechanics}, 85(1):011004, 2018.

\bibitem{ronellenfitsch2019inverse}
Henrik Ronellenfitsch, Norbert Stoop, Josephine Yu, Aden Forrow, and J{\"o}rn
  Dunkel.
\newblock Inverse design of discrete mechanical metamaterials.
\newblock {\em Physical Review Materials}, 3(9):095201, 2019.

\bibitem{goh2020inverse}
Heedong Goh and Loukas~F Kallivokas.
\newblock Inverse band gap design of elastic metamaterials for p and sv wave
  control.
\newblock {\em Computer Methods in Applied Mechanics and Engineering},
  370:113263, 2020.

\bibitem{li2020designing}
Xiang Li, Shaowu Ning, Zhanli Liu, Ziming Yan, Chengcheng Luo, and Zhuo Zhuang.
\newblock {Designing phononic crystal with anticipated band gap through a deep
  learning based data-driven method}.
\newblock {\em Computer Methods in Applied Mechanics and Engineering},
  361:112737, 2020.

\bibitem{zhang2021realization}
Xiaopeng Zhang, Jian Xing, Pai Liu, Yangjun Luo, and Zhan Kang.
\newblock Realization of full and directional band gap design by non-gradient
  topology optimization in acoustic metamaterials.
\newblock {\em Extreme Mechanics Letters}, 42:101126, 2021.

\bibitem{chen2022see}
Zhi Chen, Alexander Ogren, Chiara Daraio, L~Catherine Brinson, and Cynthia
  Rudin.
\newblock How to see hidden patterns in metamaterials with interpretable
  machine learning.
\newblock {\em Extreme Mechanics Letters}, 57:101895, 2022.

\bibitem{vismara2019solar}
Robin Vismara, Nils~Odebo L{\"a}nk, Ruggero Verre, Mikael K{\"a}ll, Olindo
  Isabella, and Miro Zeman.
\newblock Solar harvesting based on perfect absorbing all-dielectric
  nanoresonators on a mirror.
\newblock {\em Optics Express}, 27(16):A967--A980, 2019.

\bibitem{cheng2021plasmonic}
Yongzhi Cheng, Fu~Chen, and Hui Luo.
\newblock Plasmonic chiral metasurface absorber based on bilayer fourfold
  twisted semicircle nanostructure at optical frequency.
\newblock {\em Nanoscale Research Letters}, 16:1--9, 2021.

\bibitem{liu2020compounding}
Zhaocheng Liu, Dayu Zhu, Kyu-Tae Lee, Andrew~S Kim, Lakshmi Raju, and Wenshan
  Cai.
\newblock Compounding meta-atoms into metamolecules with hybrid artificial
  intelligence techniques.
\newblock {\em Advanced Materials}, 32(6):1904790, 2020.

\bibitem{Lee2024DeepNeuralOperator}
Doksoo Lee, Lu~Zhang, Yue Yu, and Wei Chen.
\newblock Deep neural operator enabled concurrent multitask design for
  multifunctional metamaterials under heterogeneous fields.
\newblock {\em Advanced Optical Materials}, 12(15):2303087, 2024.

\bibitem{inampudi2018neural}
Sandeep Inampudi and Hossein Mosallaei.
\newblock Neural network based design of metagratings.
\newblock {\em Applied Physics Letters}, 112(24), 2018.

\bibitem{zhelyeznyakov2021deep}
Maksym~V Zhelyeznyakov, Steve Brunton, and Arka Majumdar.
\newblock Deep learning to accelerate scatterer-to-field mapping for inverse
  design of dielectric metasurfaces.
\newblock {\em ACS Photonics}, 8(2):481--488, 2021.

\bibitem{liu2018training}
Dianjing Liu, Yixuan Tan, Erfan Khoram, and Zongfu Yu.
\newblock Training deep neural networks for the inverse design of nanophotonic
  structures.
\newblock {\em Acs Photonics}, 5(4):1365--1369, 2018.

\bibitem{ma2019probabilistic}
Wei Ma, Feng Cheng, Yihao Xu, Qinlong Wen, and Yongmin Liu.
\newblock Probabilistic representation and inverse design of metamaterials
  based on a deep generative model with semi-supervised learning strategy.
\newblock {\em Advanced Materials}, 31(35):1901111, 2019.

\bibitem{An2021MultifunctionalNetwork}
Sensong An, Bowen Zheng, Hong Tang, Mikhail~Y. Shalaginov, Li~Zhou, Hang Li,
  Myungkoo Kang, Kathleen~A. Richardson, Tian Gu, Juejun Hu, Clayton Fowler,
  and Hualiang Zhang.
\newblock {Multifunctional Metasurface Design with a Generative Adversarial
  Network}.
\newblock {\em Advanced Optical Materials}, 9(5):1--10, 2021.

\bibitem{Malkiel2018PlasmonicLearning}
Itzik Malkiel, Michael Mrejen, Achiya Nagler, Uri Arieli, Lior Wolf, and Haim
  Suchowski.
\newblock {Plasmonic nanostructure design and characterization via Deep
  Learning}.
\newblock {\em Light: Science and Applications}, 7(1), 2018.

\bibitem{lee2023data}
Doksoo Lee, Wei Chen, Liwei Wang, Yu-Chin Chan, and Wei Chen.
\newblock Data-driven design for metamaterials and multiscale systems: A
  review.
\newblock {\em Advanced Materials}, 36(8):2305254, 2024.

\bibitem{an2019deep}
Sensong An, Clayton Fowler, Bowen Zheng, Mikhail~Y Shalaginov, Hong Tang, Hang
  Li, Li~Zhou, Jun Ding, Anuradha~Murthy Agarwal, Clara Rivero-Baleine, et~al.
\newblock A deep learning approach for objective-driven all-dielectric
  metasurface design.
\newblock {\em ACS Photonics}, 6(12):3196--3207, 2019.

\bibitem{kumar2020inverse}
Siddhant Kumar, Stephanie Tan, Li~Zheng, and Dennis~M Kochmann.
\newblock Inverse-designed spinodoid metamaterials.
\newblock {\em npj Computational Materials}, 6(1):1--10, 2020.

\bibitem{Yeung2021MultiplexedNetworks}
Christopher Yeung, Ju~Ming Tsai, Brian King, Benjamin Pham, David Ho, Julia
  Liang, Mark~W. Knight, and Aaswath~P. Raman.
\newblock {Multiplexed supercell metasurface design and optimization with
  tandem residual networks}.
\newblock {\em Nanophotonics}, 10(3):1133--1143, 1 2021.

\bibitem{bastek2022inverting}
Jan-Hendrik Bastek, Siddhant Kumar, Bastian Telgen, Rapha{\"e}l~N Glaesener,
  and Dennis~M Kochmann.
\newblock Inverting the structure--property map of truss metamaterials by deep
  learning.
\newblock {\em Proceedings of the National Academy of Sciences},
  119(1):e2111505119, 2022.

\bibitem{xie2023deep}
Chen Xie, Haonan Li, Chenyang Cui, Haodong Lei, Yingjie Sun, Chi Zhang, Yaqiang
  Zhang, Hongxing Dong, and Long Zhang.
\newblock Deep learning assisted inverse design of metamaterial microwave
  absorber.
\newblock {\em Applied Physics Letters}, 123(18), 2023.

\bibitem{mahesh2024deep}
K~Mahesh, S~Kumar Ranjith, and RS~Mini.
\newblock A deep autoencoder based approach for the inverse design of an
  acoustic-absorber.
\newblock {\em Engineering with Computers}, 40(1):279--300, 2024.

\bibitem{ha2023rapid}
Chan~Soo Ha, Desheng Yao, Zhenpeng Xu, Chenang Liu, Han Liu, Daniel Elkins,
  Matthew Kile, Vikram Deshpande, Zhenyu Kong, Mathieu Bauchy, et~al.
\newblock Rapid inverse design of metamaterials based on prescribed mechanical
  behavior through machine learning.
\newblock {\em Nature Communications}, 14(1):5765, 2023.

\bibitem{wang2023demand}
Ze-Wei Wang, An~Chen, Zi-Xiang Xu, Jing Yang, Bin Liang, and Jian-Chun Cheng.
\newblock On-demand inverse design of acoustic metamaterials using
  probabilistic generation network.
\newblock {\em Science China Physics, Mechanics \& Astronomy}, 66(2):224311,
  2023.

\bibitem{jiang2019free}
Jiaqi Jiang, David Sell, Stephan Hoyer, Jason Hickey, Jianji Yang, and
  Jonathan~A Fan.
\newblock Free-form diffractive metagrating design based on generative
  adversarial networks.
\newblock {\em ACS nano}, 13(8):8872--8878, 2019.

\bibitem{so2019designing}
Sunae So and Junsuk Rho.
\newblock Designing nanophotonic structures using conditional deep
  convolutional generative adversarial networks.
\newblock {\em Nanophotonics}, 8(7):1255--1261, 2019.

\bibitem{wen2020robust}
Fufang Wen, Jiaqi Jiang, and Jonathan~A Fan.
\newblock Robust freeform metasurface design based on progressively growing
  generative networks.
\newblock {\em ACS Photonics}, 7(8):2098--2104, 2020.

\bibitem{gurbuz2021generative}
Caglar Gurbuz, Felix Kronowetter, Christoph Dietz, Martin Eser, Jonas Schmid,
  and Steffen Marburg.
\newblock Generative adversarial networks for the design of acoustic
  metamaterials.
\newblock {\em The Journal of the Acoustical Society of America},
  149(2):1162--1174, 2021.

\bibitem{ma2022pushing}
Wei Ma, Yihao Xu, Bo~Xiong, Lin Deng, Ru-Wen Peng, Mu~Wang, and Yongmin Liu.
\newblock Pushing the limits of functionality-multiplexing capability in
  metasurface design based on statistical machine learning.
\newblock {\em Advanced Materials}, 34(16):2110022, 2022.

\bibitem{lei2024dynamic}
Zhi-Dan Lei, Yi-Duo Xu, Cheng Lei, Yan Zhao, and Du~Wang.
\newblock Dynamic multifunctional metasurfaces: an inverse design deep learning
  approach.
\newblock {\em Photonics Research}, 12(1):123--133, 2024.

\bibitem{lew2023single}
Andrew~J Lew and Markus~J Buehler.
\newblock Single-shot forward and inverse hierarchical architected materials
  design for nonlinear mechanical properties using an attention-diffusion
  model.
\newblock {\em Materials Today}, 64:10--20, 2023.

\bibitem{zhang2023diffusion}
Zezhou Zhang, Chuanchuan Yang, Yifeng Qin, Hao Feng, Jiqiang Feng, and Hongbin
  Li.
\newblock Diffusion probabilistic model based accurate and
  high-degree-of-freedom metasurface inverse design.
\newblock {\em Nanophotonics}, 12(20):3871--3881, 2023.

\bibitem{bastek2023inverse}
Jan-Hendrik Bastek and Dennis~M Kochmann.
\newblock Inverse design of nonlinear mechanical metamaterials via video
  denoising diffusion models.
\newblock {\em Nature Machine Intelligence}, 5(12):1466--1475, 2023.

\bibitem{so2023revisiting}
Sunae So, Jungho Mun, Junghyun Park, and Junsuk Rho.
\newblock Revisiting the design strategies for metasurfaces: fundamental
  physics, optimization, and beyond.
\newblock {\em Advanced Materials}, 35(43):2206399, 2023.

\bibitem{zheng2023deep}
Xiaoyang Zheng, Xubo Zhang, Ta-Te Chen, and Ikumu Watanabe.
\newblock Deep learning in mechanical metamaterials: from prediction and
  generation to inverse design.
\newblock {\em Advanced Materials}, 35(45):2302530, 2023.

\bibitem{jin2022intelligent}
Yabin Jin, Liangshu He, Zhihui Wen, Bohayra Mortazavi, Hongwei Guo, Daniel
  Torrent, Bahram Djafari-Rouhani, Timon Rabczuk, Xiaoying Zhuang, and Yan Li.
\newblock Intelligent on-demand design of phononic metamaterials.
\newblock {\em Nanophotonics}, 11(3):439--460, 2022.

\bibitem{ma2021deep}
Wei Ma, Zhaocheng Liu, Zhaxylyk~A Kudyshev, Alexandra Boltasseva, Wenshan Cai,
  and Yongmin Liu.
\newblock Deep learning for the design of photonic structures.
\newblock {\em Nature Photonics}, 15(2):77--90, 2021.

\bibitem{jiang2021deep}
Jiaqi Jiang, Mingkun Chen, and Jonathan~A Fan.
\newblock Deep neural networks for the evaluation and design of photonic
  devices.
\newblock {\em Nature Reviews Materials}, 6(8):679--700, 2021.

\bibitem{xu2023software}
Yihao Xu, Bo~Xiong, Wei Ma, and Yongmin Liu.
\newblock Software-defined nanophotonic devices and systems empowered by
  machine learning.
\newblock {\em Progress in Quantum Electronics}, 89:100469, 2023.

\bibitem{elzouka2020interpretable}
Mahmoud Elzouka, Charles Yang, Adrian Albert, Ravi~S Prasher, and Sean~D
  Lubner.
\newblock Interpretable forward and inverse design of particle spectral
  emissivity using common machine-learning models.
\newblock {\em Cell Reports Physical Science}, 1(12), 2020.

\bibitem{breiman2001random}
Leo Breiman.
\newblock Random forests.
\newblock {\em Machine learning}, 45:5--32, 2001.

\bibitem{hastings1970monte}
W.~K. Hastings.
\newblock {Monte Carlo sampling methods using Markov chains and their
  applications}.
\newblock {\em Biometrika}, 57(1):97--109, 1970.

\bibitem{BreiFrieStonOlsh84}
Leo Breiman, Jerome Friedman, Charles~J Stone, and R~A Olshen.
\newblock {\em Classification and Regression Trees}.
\newblock Chapman and Hall/CRC, 1984.

\bibitem{chawla2002smote}
Nitesh~V Chawla, Kevin~W Bowyer, Lawrence~O Hall, and W~Philip Kegelmeyer.
\newblock Smote: synthetic minority over-sampling technique.
\newblock {\em Journal of artificial intelligence research}, 16:321--357, 2002.

\bibitem{liu2020hybrid}
Zhaocheng Liu, Lakshmi Raju, Dayu Zhu, and Wenshan Cai.
\newblock A hybrid strategy for the discovery and design of photonic
  structures.
\newblock {\em IEEE Journal on Emerging and Selected Topics in Circuits and
  Systems}, 10(1):126--135, 2020.

\bibitem{garland2021pragmatic}
Anthony~P Garland, Benjamin~C White, Scott~C Jensen, and Brad~L Boyce.
\newblock Pragmatic generative optimization of novel structural lattice
  metamaterials with machine learning.
\newblock {\em Materials \& Design}, 203:109632, 2021.

\bibitem{zhang2021genetic}
Junming Zhang, Guowu Wang, Tao Wang, and Fashen Li.
\newblock Genetic algorithms to automate the design of metasurfaces for
  absorption bandwidth broadening.
\newblock {\em ACS Applied Materials \& Interfaces}, 13(6):7792--7800, 2021.

\bibitem{shen2022nature}
Sabrina Chin-yun Shen and Markus~J Buehler.
\newblock Nature-inspired architected materials using unsupervised deep
  learning.
\newblock {\em Communications Engineering}, 1(1):37, 2022.

\bibitem{lee2022generative}
Sangryun Lee, Zhizhou Zhang, and Grace~X Gu.
\newblock Generative machine learning algorithm for lattice structures with
  superior mechanical properties.
\newblock {\em Materials Horizons}, 9(3):952--960, 2022.

\bibitem{mitchell1998introduction}
Melanie Mitchell.
\newblock {\em An introduction to genetic algorithms}.
\newblock MIT press, 1998.

\bibitem{yu2011light}
Nanfang Yu, Patrice Genevet, Mikhail~A Kats, Francesco Aieta, Jean-Philippe
  Tetienne, Federico Capasso, and Zeno Gaburro.
\newblock Light propagation with phase discontinuities: generalized laws of
  reflection and refraction.
\newblock {\em science}, 334(6054):333--337, 2011.

\bibitem{kildishev2013planar}
Alexander~V Kildishev, Alexandra Boltasseva, and Vladimir~M Shalaev.
\newblock Planar photonics with metasurfaces.
\newblock {\em Science}, 339(6125):1232009, 2013.

\bibitem{cui2014coding}
Tie~Jun Cui, Mei~Qing Qi, Xiang Wan, Jie Zhao, and Qiang Cheng.
\newblock Coding metamaterials, digital metamaterials and programmable
  metamaterials.
\newblock {\em Light: science \& applications}, 3(10):e218--e218, 2014.

\bibitem{bukhari2019metasurfaces}
Syed~S Bukhari, J~Yiannis Vardaxoglou, and William Whittow.
\newblock A metasurfaces review: Definitions and applications.
\newblock {\em Applied Sciences}, 9(13):2727, 2019.

\bibitem{hu2021review}
Jie Hu, Sankhyabrata Bandyopadhyay, Yu-hui Liu, and Li-yang Shao.
\newblock A review on metasurface: from principle to smart metadevices.
\newblock {\em Frontiers in Physics}, 8:586087, 2021.

\bibitem{landy2008perfect}
N~I{\^e} Landy, S~Sajuyigbe, Jack~J Mock, David~R Smith, and Willie~J Padilla.
\newblock Perfect metamaterial absorber.
\newblock {\em Physical review letters}, 100(20):207402, 2008.

\bibitem{liu2010infrared}
Xianliang Liu, Tatiana Starr, Anthony~F Starr, and Willie~J Padilla.
\newblock Infrared spatial and frequency selective metamaterial with near-unity
  absorbance.
\newblock {\em Physical review letters}, 104(20):207403, 2010.

\bibitem{hao2010high}
Jiaming Hao, Jing Wang, Xianliang Liu, Willie~J Padilla, Lei Zhou, and Min Qiu.
\newblock High performance optical absorber based on a plasmonic metamaterial.
\newblock {\em Applied Physics Letters}, 96(25), 2010.

\bibitem{watts2012metamaterial}
Claire~M Watts, Xianliang Liu, and Willie~J Padilla.
\newblock Metamaterial electromagnetic wave absorbers.
\newblock {\em Advanced materials}, 24(23):OP98--OP120, 2012.

\bibitem{cui2014plasmonic}
Yanxia Cui, Yingran He, Yi~Jin, Fei Ding, Liu Yang, Yuqian Ye, Shoumin Zhong,
  Yinyue Lin, and Sailing He.
\newblock Plasmonic and metamaterial structures as electromagnetic absorbers.
\newblock {\em Laser \& Photonics Reviews}, 8(4):495--520, 2014.

\bibitem{liu2017experimental}
Xinyu Liu, Kebin Fan, Ilya~V Shadrivov, and Willie~J Padilla.
\newblock Experimental realization of a terahertz all-dielectric metasurface
  absorber.
\newblock {\em Optics express}, 25(1):191--201, 2017.

\bibitem{lee2018complete}
Gun-Yeal Lee, Gwanho Yoon, Seung-Yeol Lee, Hansik Yun, Jaebum Cho, Kyookeun
  Lee, Hwi Kim, Junsuk Rho, and Byoungho Lee.
\newblock Complete amplitude and phase control of light using broadband
  holographic metasurfaces.
\newblock {\em Nanoscale}, 10(9):4237--4245, 2018.

\bibitem{zhou2018multilayer}
You Zhou, Ivan~I Kravchenko, Hao Wang, J~Ryan Nolen, Gong Gu, and Jason
  Valentine.
\newblock Multilayer noninteracting dielectric metasurfaces for multiwavelength
  metaoptics.
\newblock {\em Nano letters}, 18(12):7529--7537, 2018.

\bibitem{li2018wideband}
You Li, Qunsheng Cao, and Yi~Wang.
\newblock A wideband multifunctional multilayer switchable linear polarization
  metasurface.
\newblock {\em IEEE Antennas and Wireless Propagation Letters},
  17(7):1314--1318, 2018.

\bibitem{marino2021harmonic}
Giuseppe Marino, Davide Rocco, Carlo Gigli, Gr{\'e}goire Beaudoin, Konstantinos
  Pantzas, St{\'e}phan Suffit, Pascal Filloux, Isabelle Sagnes, Giuseppe Leo,
  and Costantino De~Angelis.
\newblock Harmonic generation with multi-layer dielectric metasurfaces.
\newblock {\em Nanophotonics}, 10(7):1837--1843, 2021.

\bibitem{malek2022multifunctional}
Stephanie~C Malek, Adam~C Overvig, Andrea Al{\`u}, and Nanfang Yu.
\newblock Multifunctional resonant wavefront-shaping meta-optics based on
  multilayer and multi-perturbation nonlocal metasurfaces.
\newblock {\em Light: Science \& Applications}, 11(1):246, 2022.

\bibitem{zhang2023high}
Yuanjian Zhang, Yingting Yi, Wenxin Li, Shiri Liang, Jing Ma, Shubo Cheng,
  Wenxing Yang, and Yougen Yi.
\newblock High absorptivity and ultra-wideband solar absorber based on ti-al2o3
  cross elliptical disk arrays.
\newblock {\em Coatings}, 13(3):531, 2023.

\bibitem{zhang2020bayesian}
Yichi Zhang, Daniel~W Apley, and Wei Chen.
\newblock Bayesian optimization for materials design with mixed quantitative
  and qualitative variables.
\newblock {\em Scientific reports}, 10(1):4924, 2020.

\bibitem{ma2020vaem}
Chao Ma, Sebastian Tschiatschek, Richard Turner, Jos{\'e}~Miguel
  Hern{\'a}ndez-Lobato, and Cheng Zhang.
\newblock Vaem: a deep generative model for heterogeneous mixed type data.
\newblock {\em Advances in Neural Information Processing Systems},
  33:11237--11247, 2020.

\bibitem{xu2018synthesizing}
Lei Xu and Kalyan Veeramachaneni.
\newblock Synthesizing tabular data using generative adversarial networks.
\newblock {\em arXiv preprint arXiv:1811.11264}, 2018.

\bibitem{liu2018generative}
Zhaocheng Liu, Dayu Zhu, Sean~P Rodrigues, Kyu-Tae Lee, and Wenshan Cai.
\newblock Generative model for the inverse design of metasurfaces.
\newblock {\em Nano letters}, 18(10):6570--6576, 2018.

\bibitem{vlassis2023denoising}
Nikolaos~N Vlassis and WaiChing Sun.
\newblock Denoising diffusion algorithm for inverse design of microstructures
  with fine-tuned nonlinear material properties.
\newblock {\em Computer Methods in Applied Mechanics and Engineering},
  413:116126, 2023.

\bibitem{li2022digital}
Weichen Li, Fengwen Wang, Ole Sigmund, and Xiaojia~Shelly Zhang.
\newblock Digital synthesis of free-form multimaterial structures for
  realization of arbitrary programmed mechanical responses.
\newblock {\em Proceedings of the National Academy of Sciences},
  119(10):e2120563119, 2022.

\bibitem{lin2023mechanical}
Xin Lin, Fei Pan, Yong Ma, Yuling Wei, Kang Yang, Zihong Wu, Juan Guan, Bin
  Ding, Bin Liu, Jinwu Xiang, et~al.
\newblock Mechanical fourier transform for programmable metamaterials.
\newblock {\em Proceedings of the National Academy of Sciences},
  120(37):e2305380120, 2023.

\bibitem{bossart2021oligomodal}
Aleksi Bossart, David~MJ Dykstra, Jop Van~der Laan, and Corentin Coulais.
\newblock Oligomodal metamaterials with multifunctional mechanics.
\newblock {\em Proceedings of the National Academy of Sciences},
  118(21):e2018610118, 2021.

\bibitem{an2021multifunctional}
Sensong An, Bowen Zheng, Hong Tang, Mikhail~Y Shalaginov, Li~Zhou, Hang Li,
  Myungkoo Kang, Kathleen~A Richardson, Tian Gu, Juejun Hu, et~al.
\newblock Multifunctional metasurface design with a generative adversarial
  network.
\newblock {\em Advanced Optical Materials}, 9(5):2001433, 2021.

\bibitem{wang2020deep}
Liwei Wang, Yu-Chin Chan, Faez Ahmed, Zhao Liu, Ping Zhu, and Wei Chen.
\newblock Deep generative modeling for mechanistic-based learning and design of
  metamaterial systems.
\newblock {\em Computer Methods in Applied Mechanics and Engineering},
  372:113377, 2020.

\bibitem{chen2023gan}
Wei Chen, Doksoo Lee, Oluwaseyi Balogun, and Wei Chen.
\newblock Gan-duf: Hierarchical deep generative models for design under
  free-form geometric uncertainty.
\newblock {\em Journal of Mechanical Design}, 145(1):011703, 2023.

\bibitem{comsol2020}
{COMSOL AB}.
\newblock Comsol multiphysics (version 5.6).
\newblock \url{https://www.comsol.com}, 2020.
\newblock Stockholm, Sweden.

\bibitem{aspnes1983dielectric}
David~E Aspnes and AA~Studna.
\newblock Dielectric functions and optical parameters of si, ge, gap, gaas,
  gasb, inp, inas, and insb from 1.5 to 6.0 ev.
\newblock {\em Physical review B}, 27(2):985, 1983.

\end{thebibliography}

\newpage

\begin{center}
{\Large Supporting Information for \\ \LARGE Generative Inverse Design of Metamaterials with Functional Responses by Interpretable Learning}
\end{center}

\appendix
\renewcommand{\thesection}{S\arabic{section}}

\renewcommand{\thefigure}{S\arabic{figure}}
\setcounter{figure}{0}
\renewcommand{\thetable}{S\arabic{table}}
\setcounter{table}{0}

\pagenumbering{arabic}
\renewcommand{\thepage}{S-\arabic{page}}

\section{Existing Approaches on Deep Learning-Based Iteration-Free, Single-Shot Inverse Design}
\label{sec:single_shot}

\begin{figure}[h]
\centering
\includegraphics[width=\textwidth]{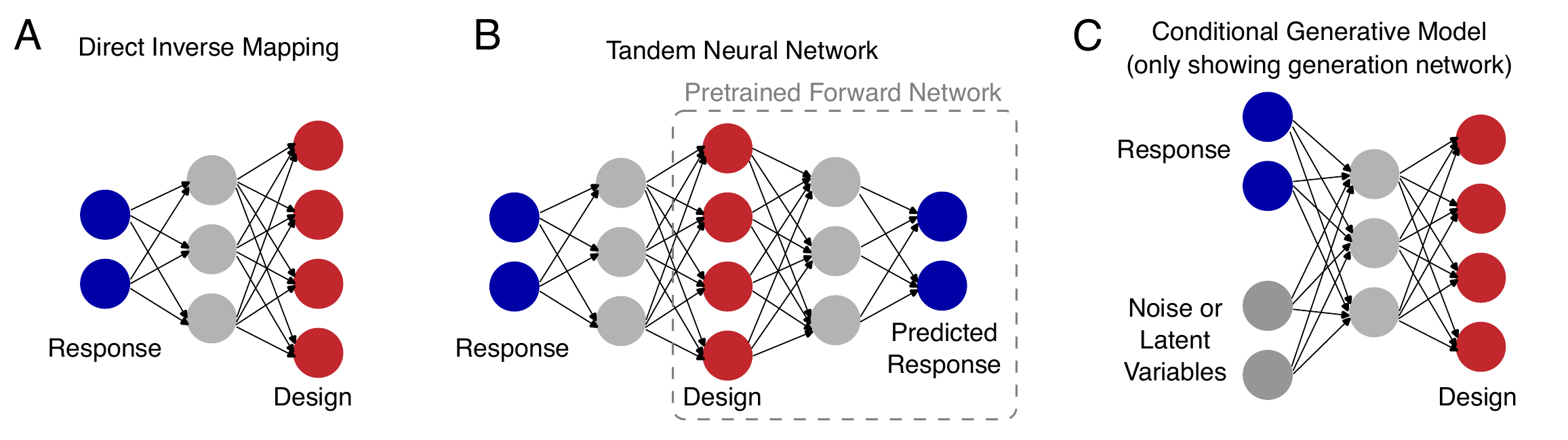}
\caption{Three major machine learning models for single-shot inverse design. Note that these are only conceptual plots and the networks can be more complicated in reality. Also, due to the diversity and complexity of conditional generative model architectures, we only show the network for design generation based on commonly used models (e.g., generators in cGANs or decoders in cVAEs), while omitting networks that are necessary for training (e.g., discriminators in cGANs or encoders in cVAEs).}
\label{fig:support_past_work}
\end{figure}

The three mainstream deep learning methods for iteration-free, single-shot metamaterial inverse design are direct inverse mapping, Tandem Neural Network (T-NN), and conditional generative models. Most prior works fall into one of the three categories or at the intersection of two categories. Here we give more details of each method. The direct inverse mapping approach (Fig.~\ref{fig:support_past_work}A) simply learns a one-to-one mapping from responses to designs. It fails to account for non-unique solutions of the inverse problem, which leads to conflicting labels (i.e., designs) for the same input (i.e., response) and hence causes training convergence issues. 

The T-NN (Fig.~\ref{fig:support_past_work}B) was designed to solve this non-uniqueness issue. Its training is split into two steps: (1)~pretraining the forward-modeling network to approximate the design-response mapping and (2)~training the cascaded network by freezing the weights of the pretrained forward-modeling network. There is no loss function that forces designs at the intermediate layer to match data (which contains conflicting instances), hence the training convergence issue is avoided. However, the original T-NNs still learn a one-to-one response-design mapping and cannot generate multiple satisfying solutions. Variants of T-NN that learn one-to-many mappings are mentioned in the main text.

The conditional generative models (Fig.~\ref{fig:support_past_work}C) are the most common approach to learning a one-to-many mapping from responses to designs. Different generative models have distinct ways of learning conditional distributions. In general, this is realized by training neural networks to transform responses and random noise (or latent variables) into designs, so that the trained network can generate a non-deterministic design solution from a given target response and randomly sampled noise, which is equivalent to sampling from a conditional distribution.

\section{Explanation of Step 2 in RIGID}

The purpose of Step 2 is to obtain the intersection of relevant design space regions for all the ranges of $s$ in a target. We are approximating this goal by simply obtaining the intersection of relevant leaves. However, some non-intersecting leaves may still have intersecting design space regions. When assigning the probability to the intersecting region of two non-intersecting leaves $A$ and $B$, we need to consider the predicted probabilities at both leaves ($P_A$ and $P_B$). Specifically, the assigned probability at this intersecting region should be $P_AP_B$, which can be small. Therefore, we adopt the simplification of only considering the intersection of relevant leaves and ignoring the intersecting regions associated with non-intersecting leaves. The results also demonstrate that this is a reasonable approximation.

\section{Test Performance of Forward Prediction}

Test data for all the design problems are designs $\mathbf{x}$ (that the random forests have never seen during training) and their corresponding qualitative behaviors $y\in\{0,1\}$ based on functional responses.

In the acoustic metamaterial design problem, we have 57 test designs, yielding 5,700 test points as the entire frequency range of dispersion relations is discretized into 100 intervals for each design. In the optical metasurface design problem, we have 52 test designs, with each functional response discretized into 33 points, which yields 1,716 test points in total. In each of the synthetic design problems, we have 20 synthetic test designs, with each synthetic response discretized into 100 points. This results in 2,000 test data points in total. Confusion matrices showing test performances are in Tables~\ref{tab:f1_acoustic}-\ref{tab:f1_test_func_sin}.

\begin{table}[H]
\caption{Confusion matrix on test data in the acoustic metamaterial design problem. The test F1 score is 0.82.}
\centering
\begin{tabular}{cc|cc}
\multicolumn{2}{c}{}
            &   \multicolumn{2}{c}{Predicted} \\
    &       &   Yes &   No              \\ 
    \cline{2-4}
\multirow{2}{*}{\rotatebox[origin=c]{90}{Actual}}
    & Yes   & 393   & 131                 \\
    & No    & 46    & 5130                \\ 
    \cline{2-4}
    \end{tabular}
\label{tab:f1_acoustic}
\end{table}

\begin{table}[H]
\caption{Confusion matrix on test data in the optical metasurface design problem. The test F1 score is 0.83.}
\centering
\begin{tabular}{cc|cc}
\multicolumn{2}{c}{}
            &   \multicolumn{2}{c}{Predicted} \\
    &       &   Yes &   No              \\ 
    \cline{2-4}
\multirow{2}{*}{\rotatebox[origin=c]{90}{Actual}}
    & Yes   & 776   & 157                 \\
    & No    & 151    & 632                \\ 
    \cline{2-4}
    \end{tabular}
\label{tab:f1_optical}
\end{table}

\begin{table}[H]
\caption{Confusion matrix on test data in the SqExp problem. The test F1 score is 0.85.}
\centering
\begin{tabular}{cc|cc}
\multicolumn{2}{c}{}
            &   \multicolumn{2}{c}{Predicted} \\
    &       &   Yes &   No              \\ 
    \cline{2-4}
\multirow{2}{*}{\rotatebox[origin=c]{90}{Actual}}
    & Yes   & 229   & 68                 \\
    & No    & 13    & 1690                \\ 
    \cline{2-4}
    \end{tabular}
\label{tab:f1_test_func_sqexp}
\end{table}

\begin{table}[H]
\caption{Confusion matrix on test data in the SupSin problem. The test F1 score is 0.86.}
\centering
\begin{tabular}{cc|cc}
\multicolumn{2}{c}{}
            &   \multicolumn{2}{c}{Predicted} \\
    &       &   Yes &   No              \\ 
    \cline{2-4}
\multirow{2}{*}{\rotatebox[origin=c]{90}{Actual}}
    & Yes   & 355   & 81                 \\
    & No    & 37    & 1527                \\ 
    \cline{2-4}
    \end{tabular}
\label{tab:f1_test_func_sin}
\end{table}

\section{Generated Designs and Corresponding Responses}

\begin{figure}
\centering
\includegraphics[width=\textwidth]{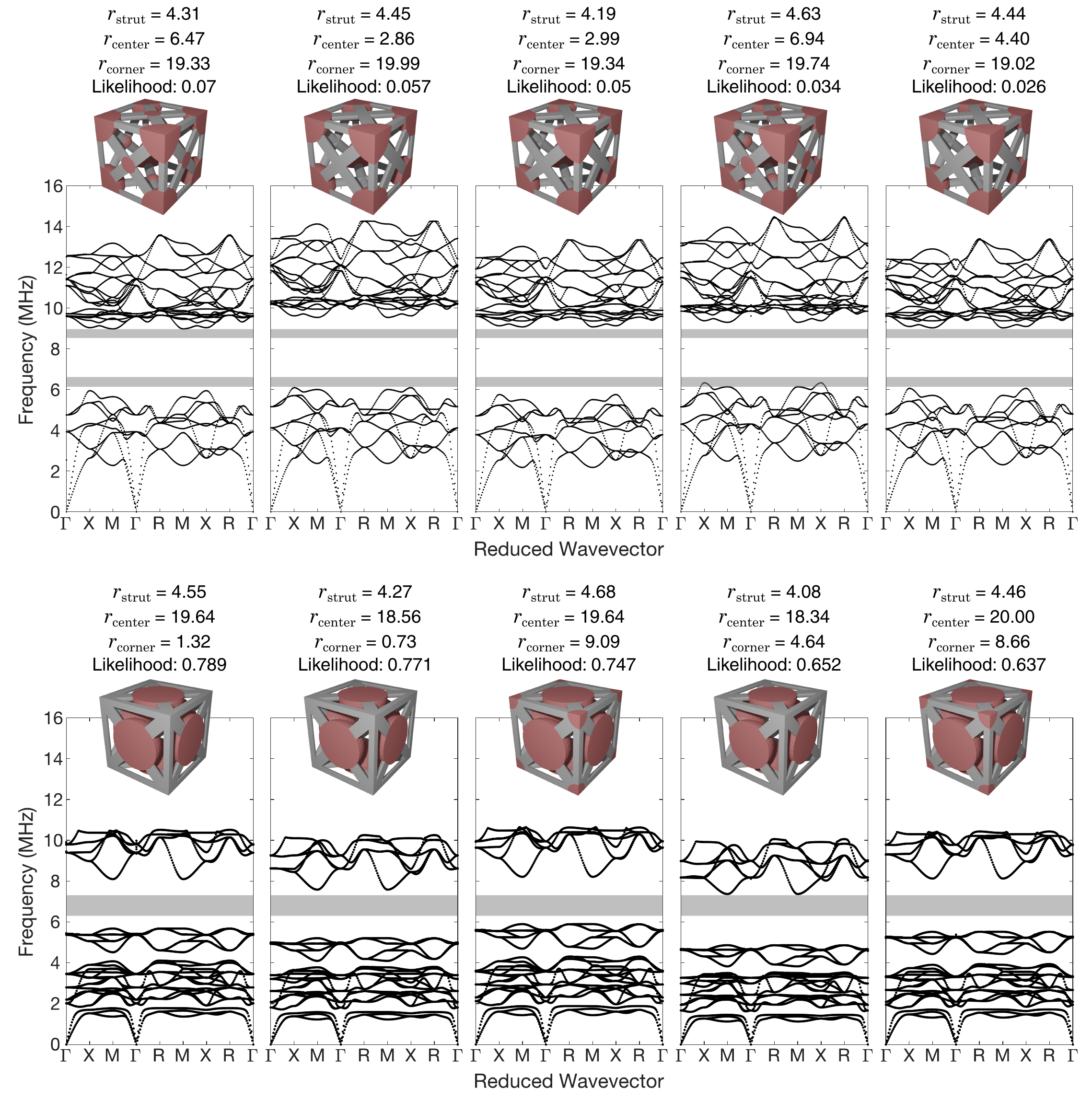}
\caption{Geometries and corresponding dispersion relations of generated acoustic metamaterial designs for bandgap targets \rom{2} and \rom{3} (marked as gray-shaded frequency regions). Each row shows five designs with the highest likelihood of satisfying a target. All the radii ($r_\text{strut}$, $r_\text{center}$, and $r_\text{corner}$) have a unit of $\mu$m.
}
\label{fig:support_acoustic_1}
\end{figure}

\begin{figure}
\centering
\includegraphics[width=\textwidth]{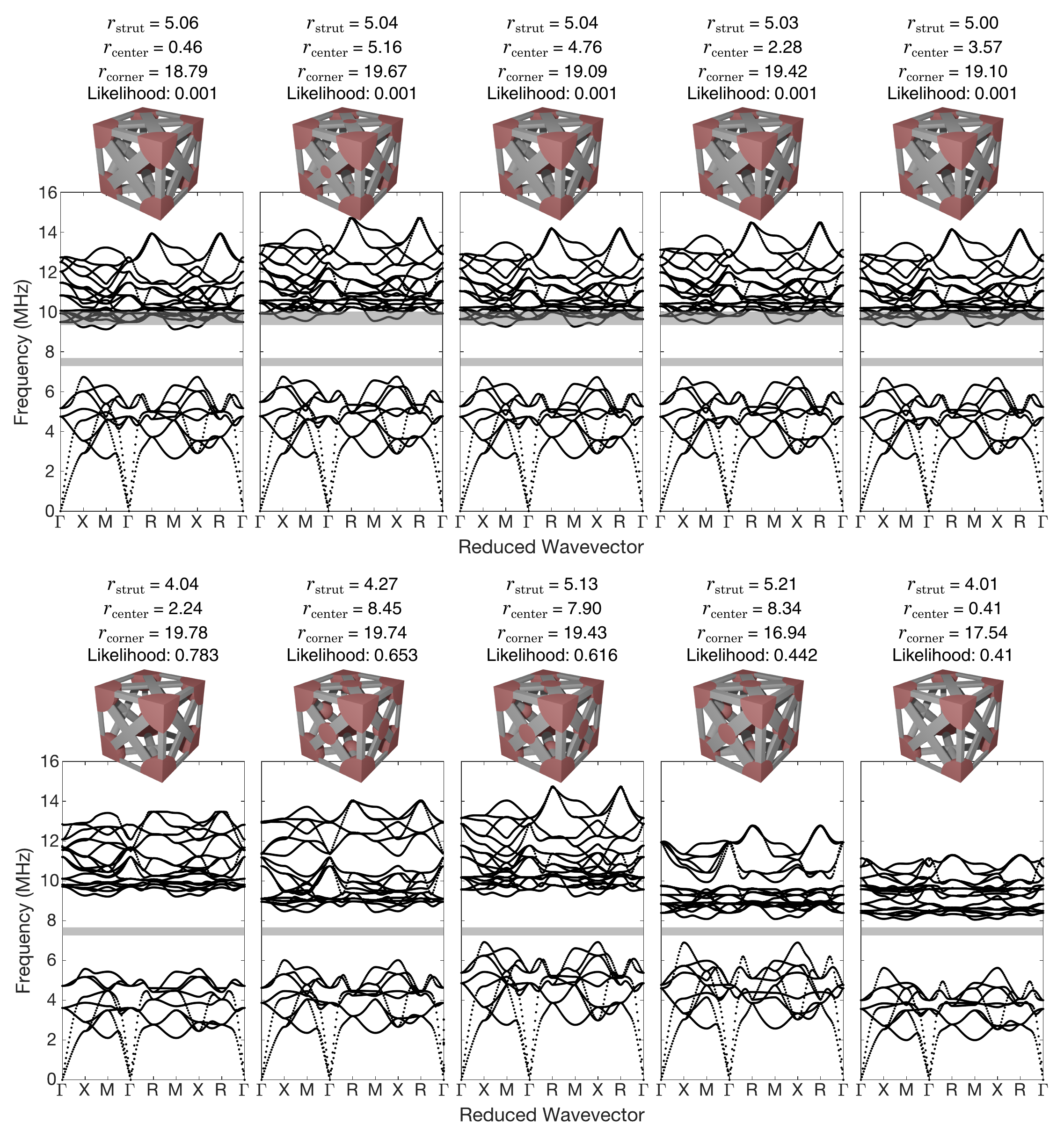}
\caption{Geometries and corresponding dispersion relations of generated acoustic metamaterial designs for bandgap targets \rom{4} and \rom{5} (marked as gray-shaded frequency regions). Each row shows five designs with the highest likelihood of satisfying a target. All the radii ($r_\text{strut}$, $r_\text{center}$, and $r_\text{corner}$) have a unit of $\mu$m.
}
\label{fig:support_acoustic_2}
\end{figure}

\begin{figure}
\centering
\includegraphics[width=\textwidth]{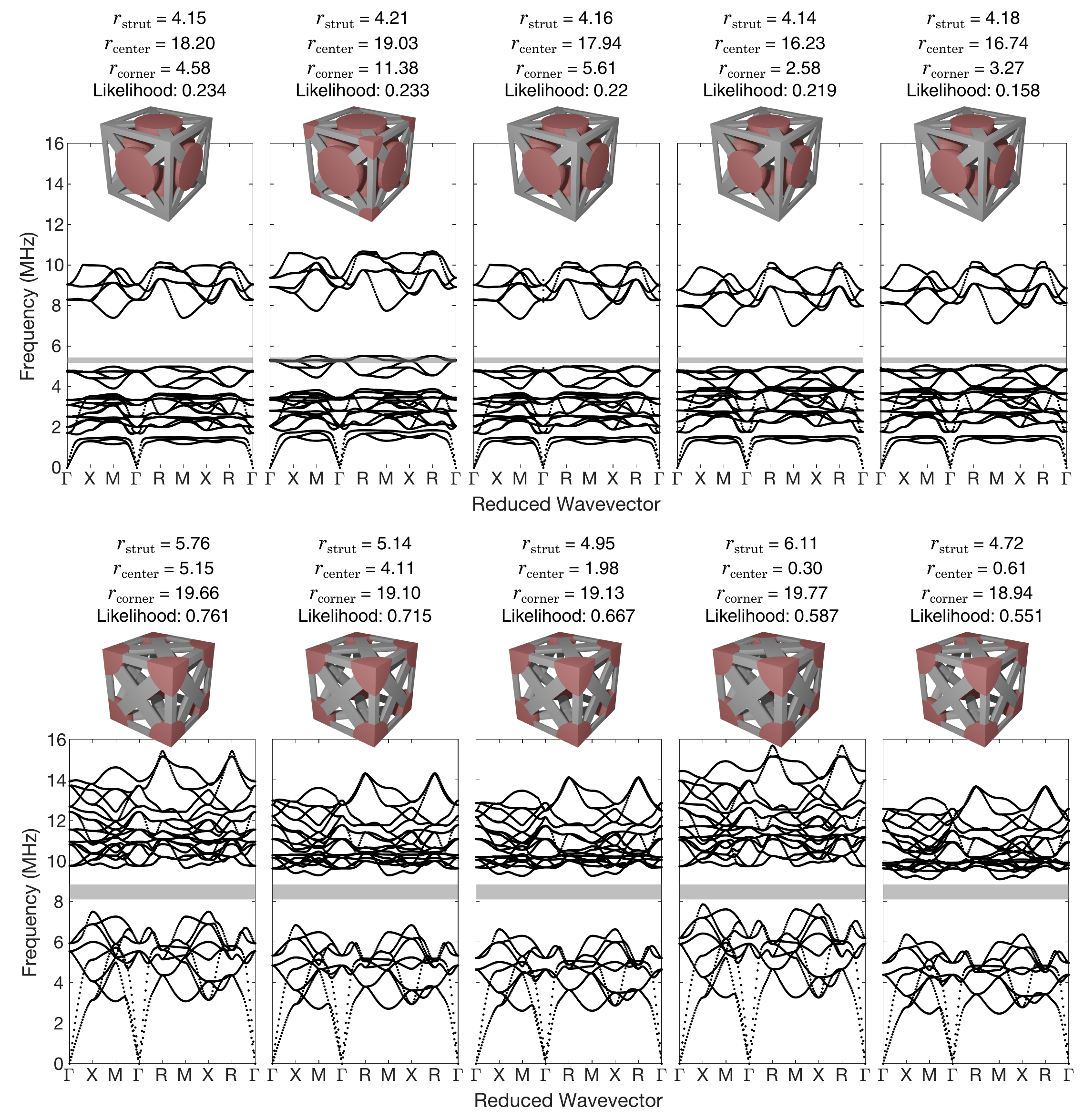}
\caption{Geometries and corresponding dispersion relations of generated acoustic metamaterial designs for bandgap targets \rom{6} and \rom{7} (marked as gray-shaded frequency regions). Each row shows five designs with the highest likelihood of satisfying a target. All the radii ($r_\text{strut}$, $r_\text{center}$, and $r_\text{corner}$) have a unit of $\mu$m.
}
\label{fig:support_acoustic_3}
\end{figure}

\begin{figure}
\centering
\includegraphics[width=\textwidth]{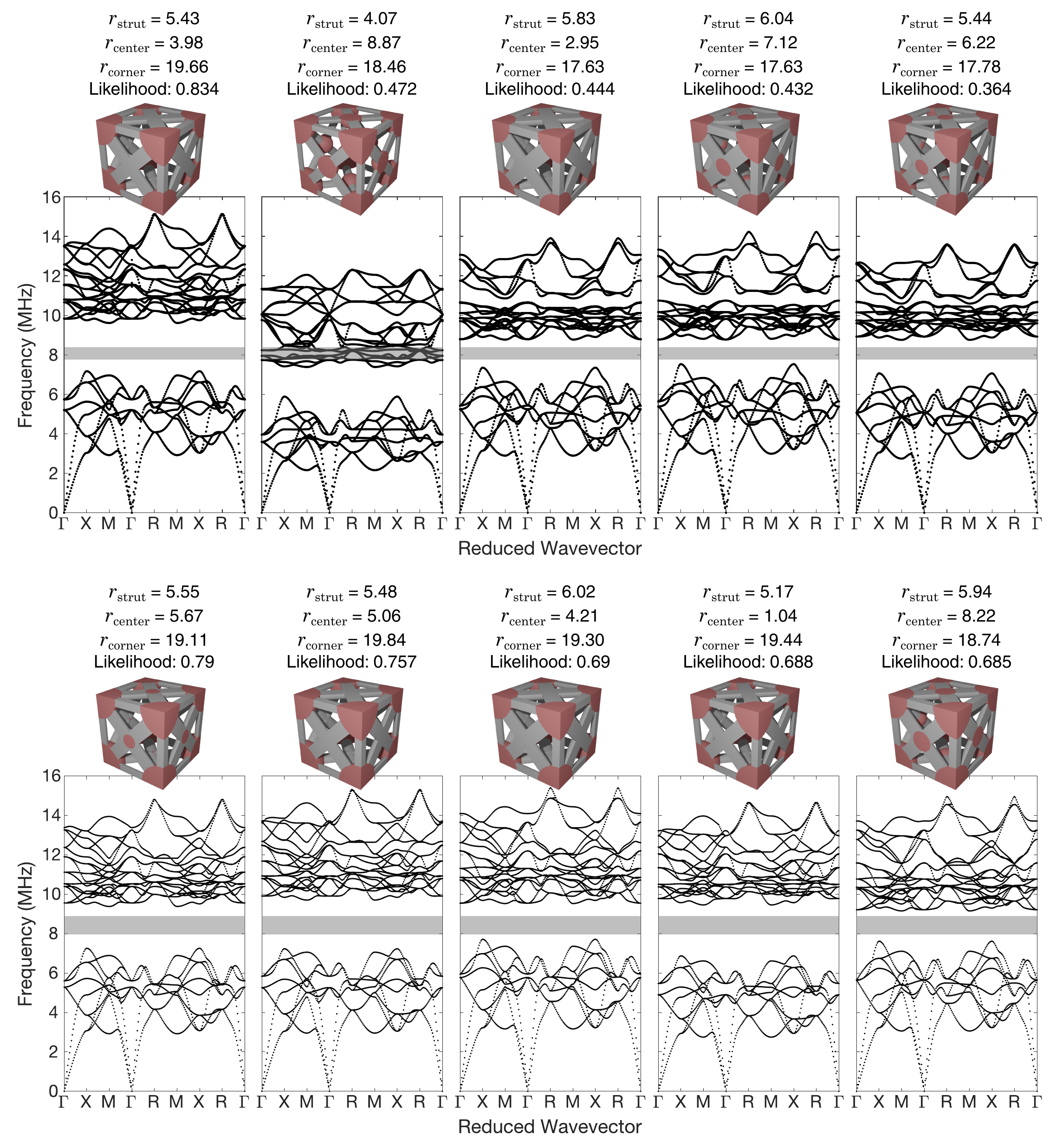}
\caption{Geometries and corresponding dispersion relations of generated acoustic metamaterial designs for bandgap targets \rom{8} and \rom{9} (marked as gray-shaded frequency regions). Each row shows five designs with the highest likelihood of satisfying a target. All the radii ($r_\text{strut}$, $r_\text{center}$, and $r_\text{corner}$) have a unit of $\mu$m.
}
\label{fig:support_acoustic_4}
\end{figure}

\begin{figure}
\centering
\includegraphics[width=\textwidth]{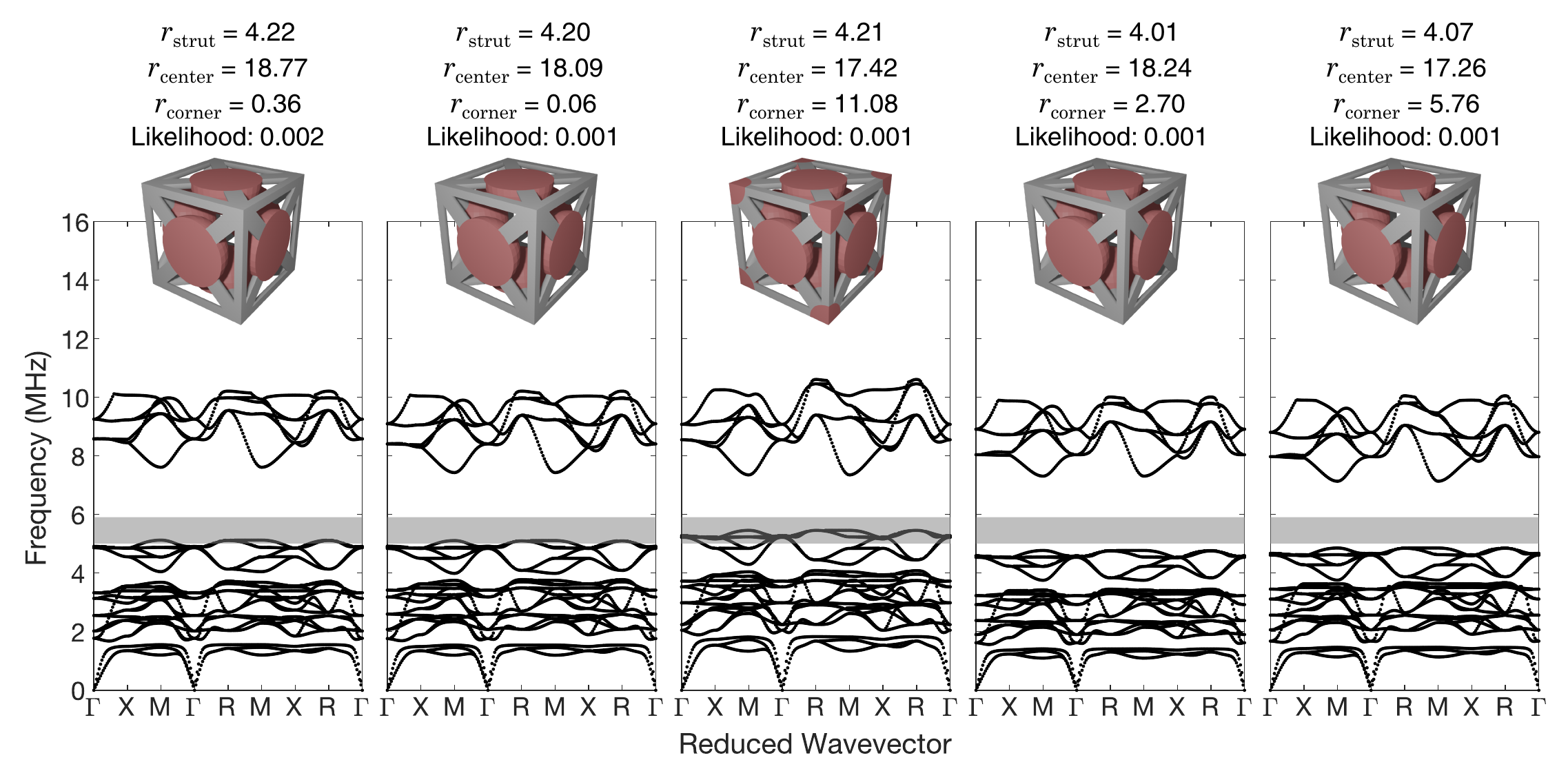}
\caption{Geometries and corresponding dispersion relations of generated acoustic metamaterial designs for bandgap target \rom{10} (marked as gray-shaded frequency regions). Each row shows five designs with the highest likelihood of satisfying a target. All the radii ($r_\text{strut}$, $r_\text{center}$, and $r_\text{corner}$) have a unit of $\mu$m.
}
\label{fig:support_acoustic_5}
\end{figure}

\begin{figure}
\centering
\includegraphics[width=\textwidth]{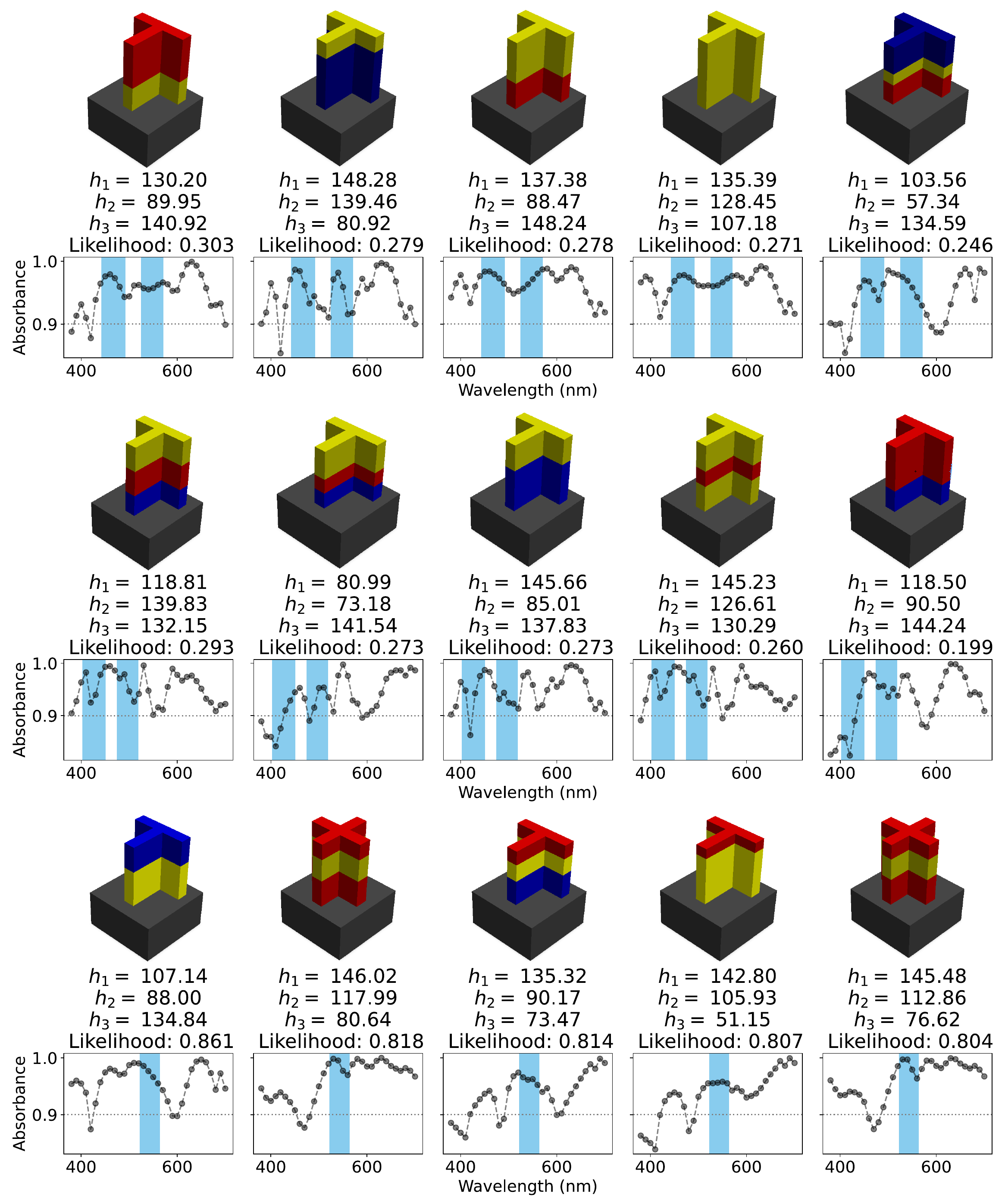}
\caption{Designs and corresponding absorbance spectra of generated optical metasurface designs for design targets \rom{2}-\rom{4} (marked as blue-shaded wavelength regions). Each row shows five designs with the highest likelihood of satisfying a target. All the layer thicknesses ($h_l, l=1, 2, 3$) have a unit of nm.
}
\label{fig:support_optical_1}
\end{figure}

\begin{figure}
\centering
\includegraphics[width=\textwidth]{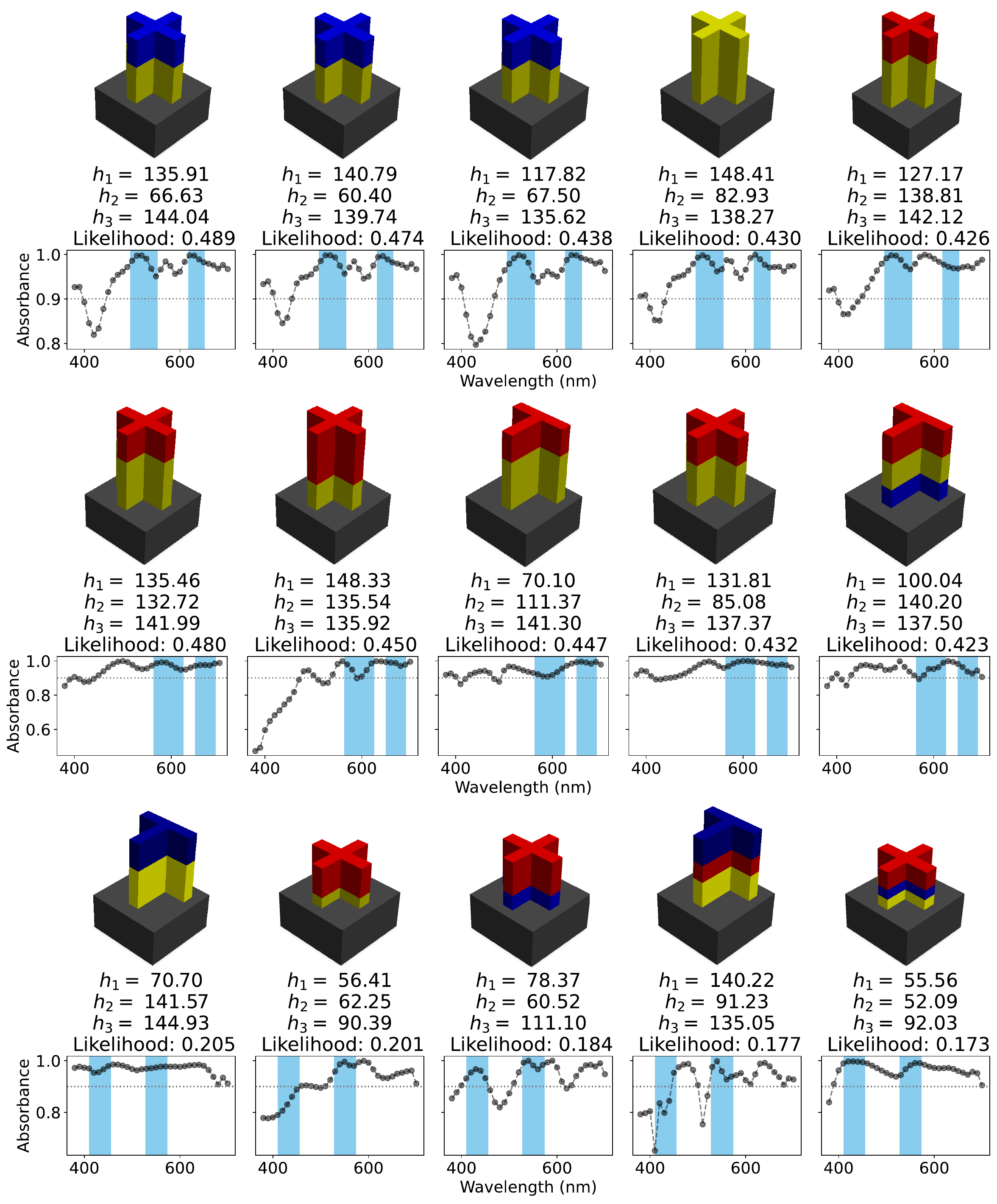}
\caption{Designs and corresponding absorbance spectra of generated optical metasurface designs for design targets \rom{5}-\rom{7} (marked as blue-shaded wavelength regions). Each row shows five designs with the highest likelihood of satisfying a target. All the layer thicknesses ($h_l, l=1, 2, 3$) have a unit of nm.
}
\label{fig:support_optical_2}
\end{figure}

\begin{figure}
\centering
\includegraphics[width=\textwidth]{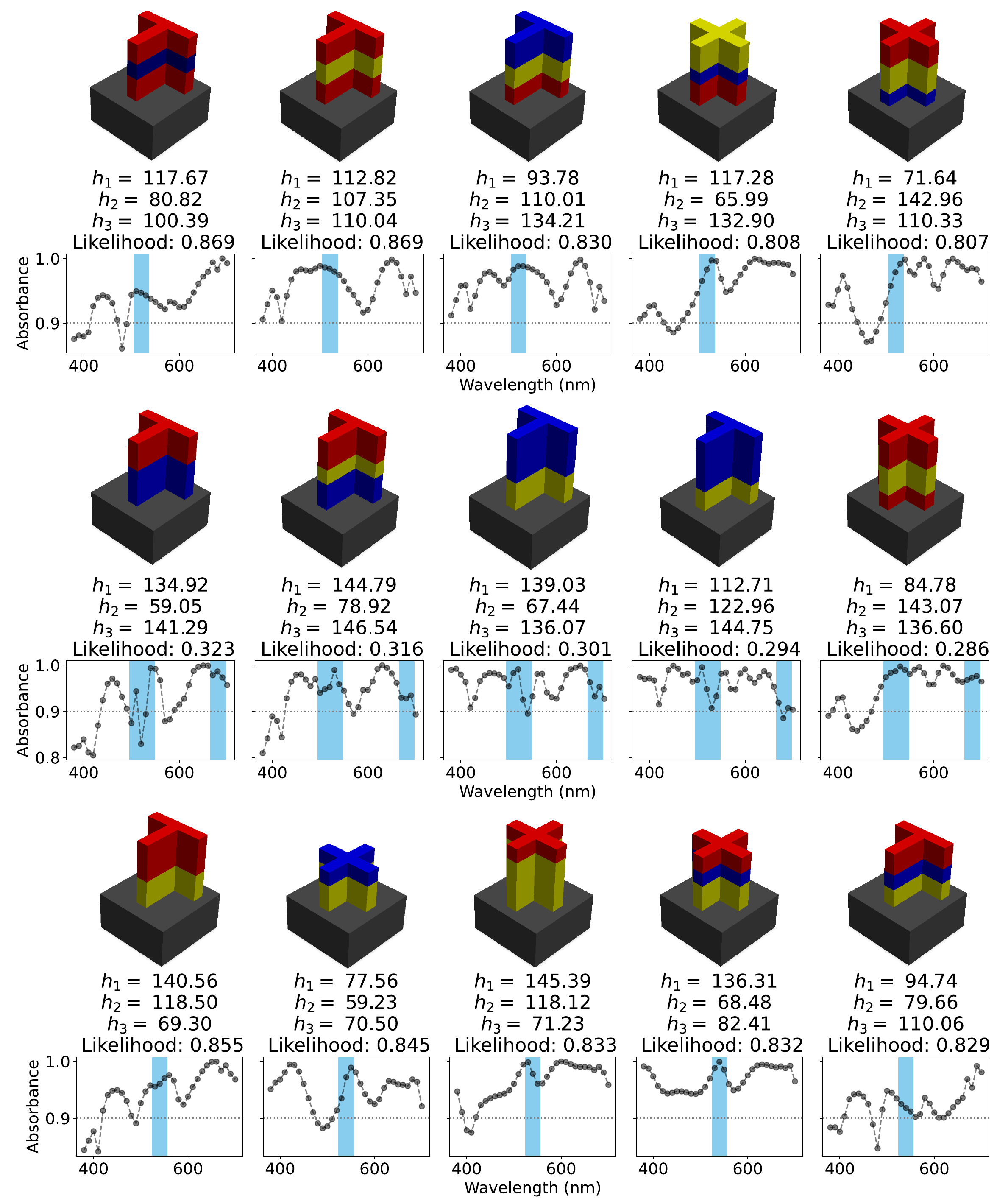}
\caption{Designs and corresponding absorbance spectra of generated optical metasurface designs for design targets \rom{8}-\rom{10} (marked as blue-shaded wavelength regions). Each row shows five designs with the highest likelihood of satisfying a target. All the layer thicknesses ($h_l, l=1, 2, 3$) have a unit of nm.
}
\label{fig:support_optical_3}
\end{figure}

In each of the acoustic and optical metamaterial design problems, we randomly created 10 design targets (i.e., frequency ranges within which bandgaps exist or wavelength ranges within which absorption is high). Given each design target, we generate multiple design solutions but only visualize five solutions having the highest likelihood values. The main text shows the five solutions for the first design target in each problem. Figures~\ref{fig:support_acoustic_1}-\ref{fig:support_optical_3} show the top solutions (ranked by likelihood) for the other design targets.

While these limited examples cannot reveal the same degree of generalizable insights as the satisfaction rate and the average score do, they still offer an intuitive glimpse into the method's performance and functionality. Most of these generated designs satisfy their corresponding targets. Some distinct designs can achieve the same target, demonstrating the capability of RIGID in solving the non-uniqueness issue of inverse design. For some targets, even the maximum likelihood of generated designs is extremely small (e.g., Target \rom{4} shown in the first row in Fig.~\ref{fig:support_acoustic_3}), indicating small likelihood values across the entire design space. This can be owing to the difficulty of achieving the target under the limit of the design space. For example, all the top five generated designs for Target \rom{4} have a likelihood of 0.001. This means that the model believes satisfying the target is almost impossible, which may be due to the wide span of the combination of two target bandgaps. This difficulty may result from the limitation of the current design space. A larger design space can lead to a wider coverage of the response space and hence more likely to contain satisfactory solutions. Based on the estimated likelihood, we can decide whether expanding the design space (i.e., extending design space bounds or adding new design freedom) is necessary to discover satisfying solutions.

\section{Investigation of Geometry Types in Generated Optical Metasurface Designs}

\begin{figure}
\centering
\includegraphics[width=.8\textwidth]{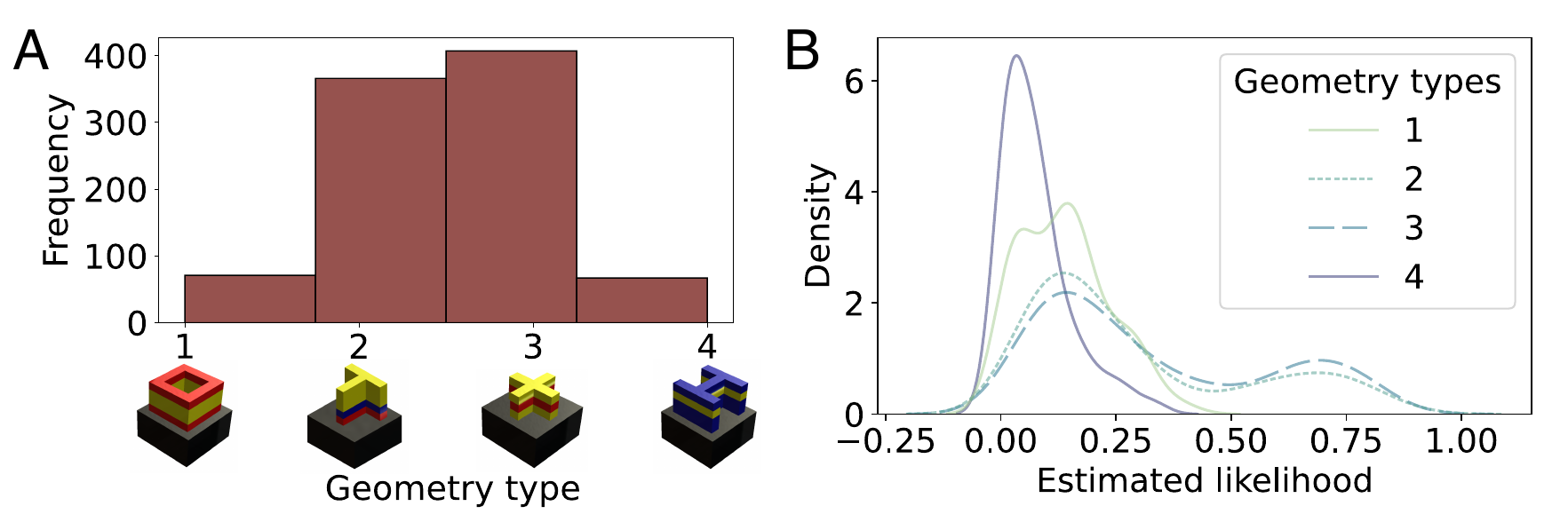}
\caption{Statistics of 1,000 generated optical metasurface designs. (A)~Histogram showing the counts of generated designs in four geometry types. (B)~KDE plot showing distributions of estimated likelihood values for different geometry types.}
\label{fig:histogram_distribution_geometry_type}
\end{figure}

\begin{figure}
\centering
\includegraphics[width=\textwidth]{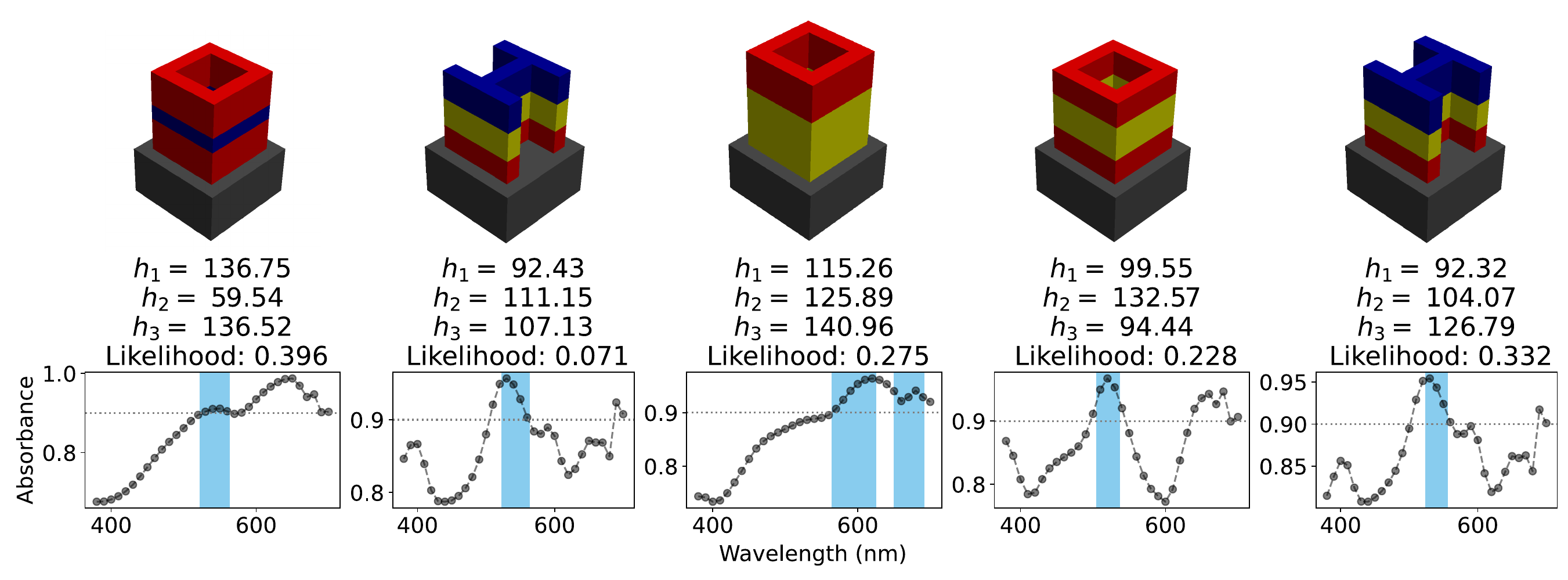}
\caption{Examples of generated designs that satisfy the targets (marked as blue-shaded wavelength regions) and belong to the rectangular ring and H-shaped geometry types. All the layer thicknesses ($h_l, l=1, 2, 3$) have a unit of nm.}
\label{fig:1and4_generated}
\end{figure}

Although the generated metasurface designs with top likelihood values in the main text and Figs.~\ref{fig:support_optical_1}-\ref{fig:support_optical_3} do not contain the rectangular ring ($c=1$) and the H-shaped ($c=4$) types, both types were generated in our inverse design processes. Based on the histogram of 1,000 generated designs from all four geometry types (Fig.~\ref{fig:histogram_distribution_geometry_type}A), there is a relatively lower percentage of designs from the rectangular ring and the H-shaped types (the 1st and 4th types). The kernel density estimation (KDE) plot (Fig.~\ref{fig:histogram_distribution_geometry_type}B) shows that these two types have fewer high-likelihood designs compared to the other two types (the 2nd and 3rd types). This explains why they are absent from the examples of generated designs with top likelihood values shown in the main text and Figs.~\ref{fig:support_optical_1}-\ref{fig:support_optical_3}. Regardless of the relatively lower likelihood values, there are satisfying designs from the rectangular ring and the H-shaped types. Some examples are shown in Fig.~\ref{fig:1and4_generated}.

The low estimated likelihood can be caused by the model's lack of belief regarding the possibility of the design meeting the target, and this belief stemmed from learning from the training dataset. Upon further investigation into our optical metasurface dataset, we found that the percentages of the high absorbance region (absorbance not less than the 0.9 threshold) within the considered wavelength range for the rectangular ring and the H-shaped geometry types are 29.5\% and 18.8\%, respectively, while the percentages for the other two types are above 80\%. This may explain the model's low confidence in believing the rectangular ring and the H-shaped geometry types satisfying the target, which further led to the low likelihood.

\end{document}